\documentclass[a4paper,11pt]{article}
\pdfoutput=1 

\usepackage{jcappub} 

\usepackage[T1]{fontenc} 

\title{H-ATLAS/GAMA: Magnification Bias Tomography. Astrophysical constraints above $\sim1$ arcmin.}

\author[a]{J. Gonz\'{a}lez-Nuevo,}
\author[b,c,d]{A. Lapi,}
\author[a]{L. Bonavera,}
\author[b]{L. Danese,}
\author[e]{G. de Zotti,}
\author[f]{M. Negrello,}
\author[g]{N. Bourne,}
\author[h]{A. Cooray,}
\author[f,g]{L. Dunne,}
\author[i]{S. Dye,}
\author[f]{S. Eales,}
\author[i,j]{C. Furlanetto,}
\author[k,g]{R. J. Ivison,}
\author[m]{J. Loveday}
\author[f,g]{S. Maddox,}
\author[f]{M. W. L. Smith,}
\author[f]{and E. Valiante,}

\affiliation[a]{Departamento de F\'{i}sica, Universidad de Oviedo, C. Federico Garc\'{i}a Lorca 18, E-33007 Oviedo, Spain}
\affiliation[b]{SISSA, Via Bonomea 265, I-34136 Trieste, Italy}
\affiliation[c]{INFN-Sezione di Trieste, via Valerio 2, 34127 Trieste, Italy}
\affiliation[d]{INAF-Osservatorio Astronomico di Trieste, via Tiepolo 11, 34131 Trieste, Italy}
\affiliation[e]{INAF, Osservatorio Astronomico di Padova, Vicolo Osservatorio 5, I-35122 Padova, Italy}
\affiliation[f]{School of Physics and Astronomy, Cardiff University, The Parade, Cardiff CF24 3AA, UK}
\affiliation[g]{Institute for Astronomy, University of Edinburgh, Royal Observatory, Blackford Hill, Edinburgh, EH9 3HJ, UK}
\affiliation[h]{Department of Physics and Astronomy, University of California, Irvine, CA, 92697, USA}
\affiliation[i]{School of Physics and Astronomy, Nottingham University, University Park, Nottingham, NG7 2RD, UK}
\affiliation[j]{Astronomy Department IF-UFRGS, Av. Bento Gonalves 9500, Agronomia PO Box 15051, 91501-970, Porto Alegre, RS, Brazil}
\affiliation[k]{European Southern Observatory, Karl-Schwarzschild-Str. 2, D-85748 Garching, Germany}
\affiliation[m]{Astronomy Centre, University of Sussex, Falmer, Brighton BN1 9QH}

\emailAdd{gnuevo@uniovi.es}

\abstract{ An unambiguous manifestation of the magnification bias is the cross-correlation between two source samples with non-overlapping redshift distributions.
In this work we measure and study the cross-correlation signal between a foreground sample of GAMA galaxies with spectroscopic redshifts in the range $0.2<z<0.8$, and a background sample of H-ATLAS galaxies with photometric redshifts $\gtrsim1.2$. It constitutes a substantial improvement over the cross-correlation measurements made by Gonzalez-Nuevo et al. (2014) with updated catalogues and wider area (with $S/N\gtrsim 5$ below 10 arcmin and reaching $S/N\sim 20$ below 30 arcsec). The better statistics allow us to split the sample in different redshift bins and to perform a tomographic analysis (with $S/N\gtrsim 3$ below 10 arcmin and reaching $S/N\sim 15$ below 30 arcsec). Moreover, we implement a halo model to extract astrophysical information about the background galaxies and the deflectors that are producing the lensing link between the foreground (lenses) and background (sources) samples. In the case of the sources, we find typical mass values in agreement with previous studies: a minimum halo mass to host a central galaxy, $M_{min}\sim 10^{12.26} M_\odot$, and a pivot halo mass to have at least one sub-halo satellite, $M_1\sim 10^{12.84} M_\odot$. 
However, the lenses are massive galaxies or even galaxy groups/clusters, with minimum mass of $M_{min}^{lens}\sim 10^{13.06} M_\odot$. Above a mass of $M_1^{lens}\sim 10^{14.57} M_\odot$ they contain at least one additional satellite galaxy which contributes to the lensing effect. The tomographic analysis shows that, while $M_1^{lens}$ is almost redshift independent, there is a clear evolution of increase $M_{min}^{lens}$ with redshift in agreement with theoretical estimations. Finally, the halo modeling allows us to identify a strong lensing contribution to the cross-correlation for angular scales below 30 arcsec. This interpretation is supported by the results of basic but effective simulations. }

\begin{document}
\maketitle
\flushbottom

\section{Introduction}
\label{sec:intro}

The gravitational lensing effect produced by a foreground deflector magnifies the light rays coming from background sources along the line-of-sight and stretches the area of the sky region around itself. Therefore, the gravitational lensing increases the probability of those amplified background sources to be detected in a flux-limited sample. This observational bias is know as `magnification bias' and it is extensively described in the literature (see, e.g., \cite{Sch92}). Depending on the slope of the background source number counts, the magnification bias produces an excess/deficit of background sources in the proximity of foreground matter overdensities \cite{Bar93,Moe98,Scr05}. Although the strong gravitational lensing effect produces higher amplification factors, $\mu>2$, the probability of these events is very low, and therefore, most of the magnification bias should be produced by the ubiquitous weak lensing effect, $\mu<2$, caused by the more common lower density cosmic structures. As a consequence, the sensitivity regarding the magnification bias will be highly enhanced in the presence of a sample of background sources with very steep source number counts. Moreover, an unambiguous manifestation of the magnification bias is the cross-correlation between two source samples with non-overlapping redshift distributions. The occurrence of such correlations has been tested and established in several contexts (see, e.g., \cite{Scr05,Men10,Hil13,Bar01}, and references therein). 

From the analysis of the first sub-millimeter sources detected by the SCUBA experiment (see \cite{Alm05} for more details) it was observed an excess of background sources,  $\sim25$ per cent, apparently associated with dense, low-redshift structures ($z\sim0.5$). 
The simplest explanation was the cross-contamination: a larger than expected fraction of background sources lying at lower redshifts. However, it was also introduced the idea of magnification bias as a probable cause for the observed cross-correlation. \cite{Bla06} performed further analysis on the issue confirming the former explanation. In these works, it was also anticipated the necessity of deeper wide area sub-mm surveys to increase the background source density in order to observe the potential magnification bias in foreground-background cross-correlation studies.

This condition was fulfilled with the launch of the \textit{Herschel} Space Observatory \cite{Pil10}. In particular, the \textit{Herschel}\footnote{\textit{Herschel} is an ESA space observatory with science instruments provided by European-led Principal Investigator consortia and with important participation from NASA} Astrophysical Terahertz Large Area Survey (\cite{Eal10}, H-ATLAS hereafter) covered around 610 square degrees with a sensitivity better than 7.3 mJy at 250 $\mu m$, becoming ideal for this kind of studies. In fact, it was already demonstrated with the Science Demonstration Phase (SDP) data that the high redshift sub-mm sources constitute an optimal sample for gravitational lensing analysis due to their main characteristics: narrow redshift ditribution, steep source number counts and low contamination from the lenses (typically elliptical galaxies around $z\sim0.5$ with negligible emission in the sub-mm band). \cite{Neg10} proposed a very simple but effective procedure to identify strong gravitational lensing events implying an estimated rate of 0.13-0.21 strongly lensed galaxies per square degree. Later on, \cite{GN12} proposed a more complex procedure for the identification of such rare events, with an upgraded rate of 1.5 strongly lensed galaxies per square degree. These works opened up a new field of research on the (statistical) analysis of the strongly lensed galaxies identified in the sub-mm band \cite{Neg14,Neg17,Fu12,Bus12,Bus13,War13,Cal14,Nay16}. It was also extended at millimeter wavelengths with the South Pole Telescope survey (e.g. \cite{Vie13,Spi16}) and the Planck all sky surveys \cite{PIPXXVII,Can15,Har16,Nes16,Can17}.

It was during the analysis performed by \cite{GN12} that it was noticed a tendency of high-z H-ATLAS sources to appears along the same line-of-sight of foreground overdensities, similar to \cite{Alm05}. The same issue was identified by \cite{Bou14} observing an excess of red galaxies, i.e. high redshift, cross-matched with foreground galaxies. A first attempt at measuring lensing-induced cross-correlations between Herschel/SPIRE galaxies and low-z galaxies was carried out by \cite{Wan11} using only the prelimnary SDP catalogue, who found convincing evidence of the effect. With much better statistics, this potential bias was studied in detail in \cite[hereafter GN14]{GN14} by measuring the angular cross-correlation function between selected HATLAS high-z sources, $z>1.5$, and two optical samples with redshifts $0.2<z<0.6$, extracted from the Sloan Digital Sky Survey \cite[SDSS]{Ahn12}) and Galaxy and Mass Assembly \cite[GAMA]{Dri11} surveys. The observed cross-correlation function was measured with high significance, $>10\sigma$. Moreover, based on realistic simulations, it was concluded that the signal was entirely explained by a magnification bias produced by the weak lensing effect caused by low redshift cosmic structures (galaxy groups/clusters with halo masses in the range $10^{13.2}$--$10^{14.5} M_\odot$) signposted by the brightest galaxies in the optical samples.

This work constitute a step forward with respect to the GN14 results by improving the significance of the measured cross-correlation signal, by splitting the foreground sample in different redshift bins to perform a tomographic analysis and by applying a theoretical halo modeling framework to extract useful astrophysical information about the objects acting as deflectors and their evolution with redshift. The paper is structured as follows. In Section 2, we describe the theoretical background needed to apply the halo model approach to the measured signal. The selection of background and foreground samples and their characteristics is presented in Section 3. In Section 4, we describe the procedure used to perform the different types of measurements. The main results are discussed in detail in Section 5 and summarized in Section 6.

Throughout the paper, we adopt a flat $\Lambda$CDM cosmology with the best-fit cosmological parameters determined by Planck Collaboration 2015:   matter density $\Omega_m$ = 0.31, $\Omega_\lambda$ = 0.69, $\sigma_8=0.82$ and Hubble constant h = $H_0$ /100 km s$^{−1}$ Mpc$^{−1}$ = 0.68.

\section{Theoretical background} \label{sec:theo}
\subsection{Galaxy-mass correlation: Halo model formalism}

In order to interpret a foreground-background source cross-correlation signal we adopt the halo model formalism \cite{Coo02}. During the last decade it become a standard to parametrize the power spectrum of the galaxy distribution as the sum of a 2-halo term, related to the correlations between different halos that dominates at large scales, and a 1-halo term, more important at small scales, that depends on the distribution of galaxies within the same halo. Moreover, the halo model also suggest a simple parametrization of the cross-correlation between the galaxy and dark matter distributions \cite{Sel00,Guz01,Coo02}:
\begin{subequations}\label{eq:hpk}
\begin{equation}
P_\mathrm{gal-dm}(k,z) = P^{1h}_\mathrm{gal-dm}(k,z) + P^{2h}_\mathrm{gal-dm}(k,z),
\end{equation}
where
\begin{equation}
P^{1h}_\mathrm{gal-dm}(k,z) = \int{dM \frac{M \frac{dN}{dM}(z)}{\overline{\rho}}\frac{\langle N_\mathrm{gal} | M\rangle}{\overline{n}_\mathrm{gal}} \vert u_\mathrm{dm} (k|M,z)\vert \vert u_\mathrm{gal} (k|M,z)\vert^{p-1} },
\end{equation}
\begin{equation}
\begin{split}
P^{2h}_\mathrm{gal-dm}(k,z) = P^\mathrm{lin}(k,z) &\left[ \int{dM \frac{M \frac{dN}{dM}(z)}{\overline{\rho}}} b(M,z)u_\mathrm{dm} (k|M,z)\right]\times \\
&\times\left[ \int{dM \frac{dN}{dM}(z)\frac{\langle N_\mathrm{gal}|M\rangle}{\overline{n}_\mathrm{gal}} b(M,z)u_\mathrm{gal}(k|M,z)}\right].
\end{split}
\end{equation}
\end{subequations}
In these equations $\overline{\rho}$ is the background density, $\frac{dN}{dM}(z)$ is the halo mass function \cite{Set99}, $P^\mathrm{lin}(k,z)$ is the linear dark matter power spectrum, $b(M,z)$ is the linear large-scale bias, and $u_\mathrm{gal} (k|M,z)$ is the normalized Fourier transform of the galaxy density distribution within a halo, which is assumed to equal the dark matter density profile, i.e. $u_\mathrm{gal} (k|M,z)=u_\mathrm{dm} (k|M,z)$. We define halos here as overdense regions whose mean density is 200 times the mean background density of the universe according to the spherical collapse model, and we adopt the density profile of \cite[hereafter NFW]{NFW96} with the concentration parameter of \cite{Bul01}.

The mean number of galaxies is represented by $\overline{n}_\mathrm{gal}$ while $\langle N_\mathrm{gal}|M\rangle$ is the mean number of galaxies in a halo of mass $M$, where, as usual, we make the distinction between central and satellite galaxies, $N_\mathrm{gal} = N_\mathrm{cen} + N_\mathrm{sat} = 1 +N_\mathrm{sat}$.  All halos above a minimum mass $M_\mathrm{min}$ host a galaxy at their center, while any remaining galaxies are classified as satellites and are distributed in proportion to the halo mass profile (see e.g. \cite{Zen05}). Halos host satellites when their mass exceeds the $M_1$ mass, and the number of satellites is a power-law function of halo mass:
\begin{equation}\label{eq:hod}
N_\mathrm{sat}(M) = \left(\frac{M}{M_1}\right)^{\alpha_\mathrm{sat}}.
\end{equation}
These parameters define the adopted Halo Occupation Distribution (HOD). Therefore, in Eq. \ref{eq:hpk} we assume $p=1$ if the halo contains only one galaxy (it will sit at the center) and $p=2$ otherwise.

\subsection{Foreground-background source angular cross-correlation function}
The cross-correlation between the galaxy and dark matter distributions, see \eqref{eq:hpk}, can be probed through two independent methods: the weak lensing tangential shear-galaxy correlation and the foreground-background source correlation function. In this paper we focus on the second method. 
The dependence of the galaxy-mass correlation on the foreground-background source correlation arises from the weak lensing effect, affecting the source number counts of the background galaxy sample (\textit{magnification bias}), that is produced by the mass density field which is traced by the foreground galaxy sample.

Following mainly \cite{Coo02}, we can write the correlation between the foreground and background sources as:
\begin{equation}
\omega_{fb}(\theta)=\langle\delta N_f(\hat{n})\delta N_b(\hat{n}+\theta)\rangle .
\end{equation}

The foreground sources are assumed to trace the density field and based on the source clustering one can write the fluctuations in the foreground source population as
\begin{equation}
\delta N_f(\hat{n})=\int_0^{z_s} dz\, \frac{dN_f}{dz} \delta_\mathrm{gal}(\hat{n},z),
\end{equation}
with $ \frac{dN_f}{dz}$ as the \textit{unit-normalized} foreground redshift distribution and $z_s$ the source redshift.

In the case of the background sources, whose number counts can be written as $N(S)= N_0 S^{-\beta}$, we know that in the presence of lensing we have amplification and dilution effects: $N(S)=\frac{N_0}{\mu}(\frac{S}{\mu})^{-\beta}$. For weak lensing the amplification can be approximated by $\mu\simeq 1+2\kappa$, and therefore:
\begin{equation}
\begin{split}
\delta N_b(\hat{n}) &= 2(\beta-1)\kappa(\hat{n}) \\
&= 2(\beta-1)\int_0^{z_s}dz\, \mathrm{W^{lens}}(z)\delta_\mathrm{dm}(\hat{n},z),
\end{split}
\end{equation}
with
\begin{equation}
\mathrm{W^{lens}}(z)= \frac{3}{2}\frac{H^2_0}{c^2} E^2(z) \int_z^{z_s}dz'\frac{\chi(z)\chi(z'-z)}{\chi(z')}\frac{dN_b}{dz'}.
\end{equation}
Here, $\chi(z)$ is the comoving distance to redshift $z$, $E(z)=\sqrt{\Omega_M(1+z)^3+\Omega_\Lambda}$ and $\frac{dN_b}{dz}$ is the \textit{unit-normalized} background redshift distribution.

Therefore, the correlation between the foreground and background sources can be evaluated as:
\begin{equation}
\begin{split}
\omega_{fb}(\theta)&= 2(\beta-1)\int_0^{z_s} dz\, \frac{dN_f}{dz} \mathrm{W^{lens}}(z) \langle\delta_\mathrm{gal}(\hat{n},z) \delta_\mathrm{dm}(\hat{n}+\theta,z)\rangle \\
&= 2(\beta-1)\int_0^{z_s} \frac{dz}{\chi^2(z)}\, \frac{dN_f}{dz} \mathrm{W^{lens}}(z) \int_0^\infty{\frac{\ell d\ell}{2\pi}P_\mathrm{gal-dm}(\ell/\chi(z),z)J_0(\ell\theta),}
\end{split}
\end{equation}
where we have made use of the standard Limber \cite{Lim53} and flat-sky approximations (see for example \cite{Kil17} and references therein).

We can finally interpret the cross-correlation signal under the halo model parametrization taking into account that both galaxy samples trace the same dark matter distribution around redshift $z~\sim0.4$. This dark matter distribution is traced directly by the foreground galaxies while, in the case of the background sample, it is traced thanks to the weak lensing effect. In this framework, the 2-halo term corresponds to the correlation between one halo traced by the foreground galaxies and another one traced by the background sources. In a similar way, we have the 1-halo term that describes the correlation between sub-halos (traced by both samples) inside the same halo.

\section{Data}
In this section we describe the selection and details of the background and foreground samples.

\begin{figure}[tbp]
\centering 
\includegraphics[width=\textwidth]{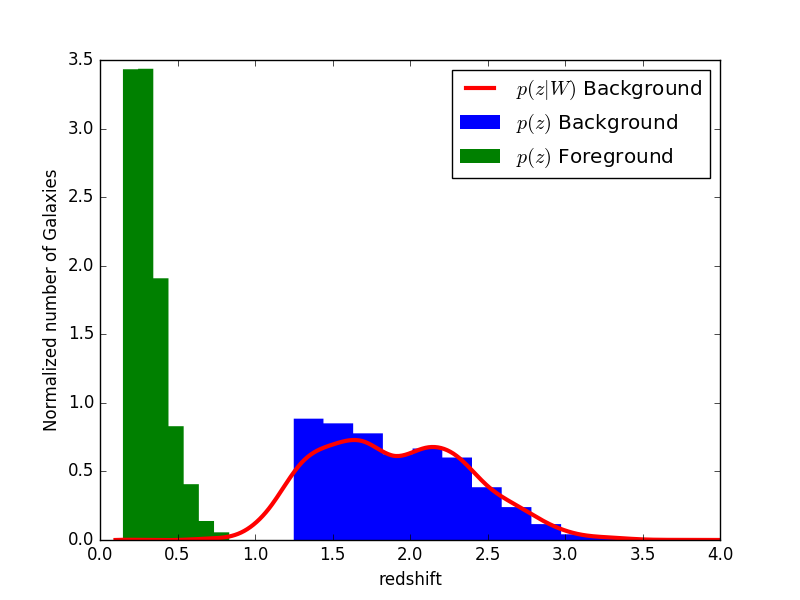}
\caption{\label{fig:zhist} Redshift distributions of the background H-ATLAS sample (blue histogram) and the foreground GAMA one (green histogram). The estimated $p(z|W)$ of the background sample taking into account the window function and the photometric redshift errors is represented as a red line. The overlap around $z\sim0.75$ amounts only $\sim 0.34$ per cent of the foreground sources absolute number in the same redshift bin.}
\end{figure}

\subsection{Background  sample}
The H-ATLAS is the largest area extragalactic survey carried out by the Herschel space observatory \cite{Pil10} covering $\sim 610\, \rm{deg}^2$ with PACS \cite{Pog10} and SPIRE \cite{Gri10} instruments between 100 and $500 \mu m$. Details of the H-ATLAS map-making, source extraction and catalogue generation can be found in \cite{Iba10,Pas11,Rig11,Val16,Bou16} and Maddox et al. (in preparation). 

We have selected our background sample from the sources detected in the three H-ATLAS equatorial fields, covering altogether  $\sim147\, \rm{deg}^2$, in addition to the part of the South Galactic Pole region (SGP) overlapped with the foreground sample ($\sim60 \rm{deg}^2$, see next subsection). Therefore, the total common area between both samples used in this work is $\sim207 \rm{deg}^2$.

Similarly to GN14, for the SGP region, we first select those sources with $S_{250\mu m} > 35$ mJy and at least $3\sigma$ detections at $350\mu m$. This catalogue is still unpublished and it was processed in the same way as the catalogues used in GN14. On the other hand, the equatorial region catalogues were recently updated and publicly delivered (see \cite{Val16,Bou16} for particular details) using an updated pipeline that change slightly the detection and statistical properties of the detected sources. For this reason, we modify the criteria for the equatorial regions in order to obtain a similar source density with both catalogues: at least $4 \sigma$ detections at $250\mu m$ ($\sim 29$ mJy) and $3\sigma$ detections at $350\mu m$. While the main selection is made based on the estimated photometric redshifts, as explain below, these criteria are introduced in order to remove sources with photometry problems. Moreover, they also help in eliminating faint local sources, that are not needed in our analysis, and therefore in speeding up the process. In fact, after applying these criteria we are left with $\sim 170000$ sources that constitute the $\sim 59$ per cent of the total number of initial sources.

Next we estimate the photometric redshifts of the selected galaxies by means of a minimum $\chi^2$ fit of a template SED to the SPIRE data (using PACS data when possible). As shown by \cite{Lap11,GN12} a good template is the SED of SMM J2135-0102 (`The Cosmic Eyelash' at $z = 2.3$; \cite{Ivi10,Swi10}). A comparison with 36 sources with spectroscopic redshifts between $0.5<z<4.5$ has shown that the use of this template does not introduce any systematic offset and has reasonably low rms error (median $\Delta z/(1 + z)\equiv  (z_{phot} - z_{spec})/(1 + z_{spec}) = −0.002$ with a dispersion of 0.115 and no outliers \cite{GN12}). Similar conclusions were obtained by \cite{Pea13}, confirming our approach. 

More recently, \cite{Ivi16} re-checked again the systematic uncertainties associated with the photometric redshift estimated using single SED templates for high-z ($z>1$) sub-mm galaxies. Considering a sample of 69 bright dusty star forming galaxies with spectroscopic redshifts determined via detections of CO using broadband spectrometers (e.g., \cite{Rie13,Wei13,Asb16,Str16}), they confirmed the impressive prediction power of this approach: `The Cosmic Eyelash' was found to be the best overall template with $\Delta z/(1 + z) = -0.07$ and a dispersion of 0.153.

Taking into account the confirmed photometric redshift prediction power of `The Cosmic Eyelash' SED template, and in order to increase the background sample statistics, we decided to include all sources with photometric redshift $z >1.2$ (at difference with GN14 that limited the background sample to $z>1.5$).  Therefore, our background sample comprises $\sim 41500$ sources in total (only $\sim 24$ per cent of the initial value) with a median redshift of $\sim 1.9$. The estimated redshift distribution of selected sources is shown in Fig. \ref{fig:zhist}. 

Finally, to allow for the effect on $dN/dz$ of random errors in photometric redshifts, we estimated the redshift distribution, $p(z|W)$ (red line in Fig \ref{fig:zhist}), of galaxies selected by our window function $W(z_{\rm ph})$ (a top-hat for $1.2 < z < 4.0$) ,  as
\begin{equation}
p(z|W) = p(z) \int dz_{\rm ph} W(z_{\rm ph})p(z_{\rm ph}|z),
\end{equation}
where $p(z)$ is the initial redshift distribution, $W=1$ for $z_{\rm ph}$ in the selected interval and $W=0$ otherwise, and $p(z_{\rm ph}|z)$ is the probability that a galaxy with a true redshift $z$ has a photometric redshift $z_{\rm ph}$ \cite{Bud03,Bia16}. The error function $p(z_{\rm ph}|z)$  is parameterized as a Gaussian distribution with zero mean and variance $(1+z)\sigma_{\Delta z/(1 + z)}$. For the dispersion we adopt the updated value of $\sigma_{\Delta z/(1 + z)} = 0.153$, found by \cite{Ivi16}.

\subsection{Foreground sample}
A simple lesson we learned from GN14 was  the importance of accurate redshift measurements of the foreground sources in any study regarding the gravitational lensing effect, even more, if we are aimed toward an analysis of tomographic measurements, as is the case of the current paper.

For these reasons, the foreground sources were drawn from a spectroscopic survey: the GAMA II \cite{Dri11, Bal10, Bal14, Lis15}. Moreover, both surveys, H-ATLAS and GAMA II, were coordinated in order to maximize the overlap. In particular both surveys observed the three equatorial regions at 9, 12 and 14.5 h (referred to as G09, G12 and G15, respectively) and GAMA II observed a portion of the SGP region surveyed by H-ATLAS. The common area covered by both surveys is around $\sim 207 \rm{deg}^2$ in size, and is surveyed down to a limit of  $r \simeq 19.8$ mag. 

For our main sample we select all GAMA II galaxies (from the catalogue designated as SpecObjv27) with reliable redshift measurements and $0.2 < z < 0.8$, although we have also studied the cross-correlation signal produced by foreground sources with $0.1 < z < 0.2$. The main sample comprises  $\sim 150000 $ galaxies in total. Their median redshift, $z_{spec,med}$ = 0.28 is significantly lower than the background sample, as shown by the green histogram in Fig. \ref{fig:zhist}. 

From Fig. \ref{fig:zhist} it is clear that taking into account the photometric errors produces a broadening of the background redshift distribution toward lower redshifts. However, it is not possible to estimate the relative importance of the overlap, seen mainly around $z\sim0.75$, due to the normalization introduced to compare both distributions. The number of background sources estimated at $z\sim0.75$ amounts only $\sim 0.34$ per cent of the foreground ones in the same redshift bin, and clearly, even lower at lower redshifts. Therefore, even taking into account the background photometric uncertainties, we can consider the possible cross-contamination (sources at lower redshift, $z<0.8$, with photometric redshifts $>1.2$) statistically negligible (see \cite{Lap11,GN12,GN14} for a more detailed discussion and section \ref{sec:xc} for additional tests on this topic).

\section{Measurements}
\subsection{Foreground/Background samples angular auto-correlation}\label{sec:ac}

\begin{figure}[tbp]
\centering 
\includegraphics[width=\textwidth]{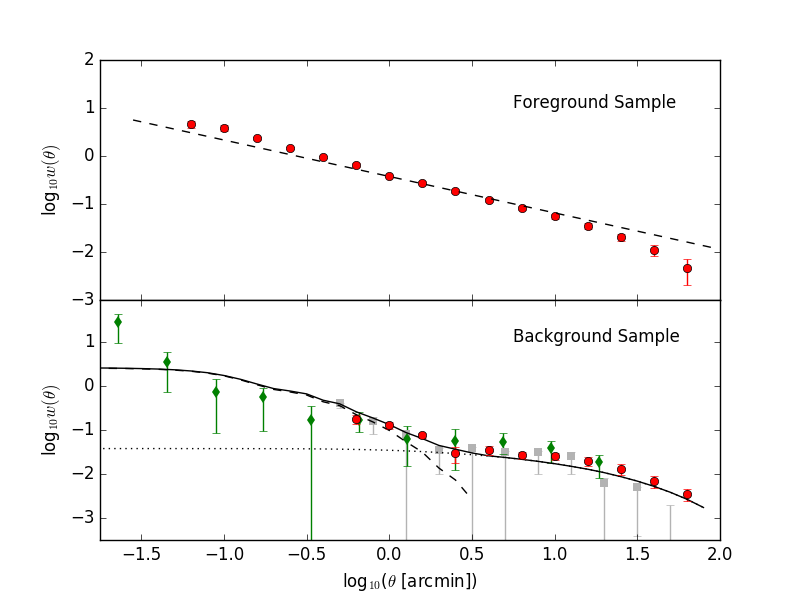}
\caption{\label{fig:ac} \textit{Top panel}: Angular auto-correlation function of the foreground sample \textit{red circles} compared with a determination of the best-fitting power law using the full SDSS catalogue for the magnitude interval $18<r<19$  (\textit{dashed line}, \cite{Con02,Wan13}). \textit{Bottom panel}: Angular auto-correlation of the background sample ($1.2 < z < 4$; \textit{red circles}) and the best halo model fit (black lines: total (\textit{solid}), 1-halo (\textit{dashed}) and 2-halo (\textit{dotted}); see text for more details). The gray squares correspond to the measurements obtained by GN14. The green diamonds are the auto-correlation measured by \cite{Che16b} for SMG between $2<z<3$.}
\end{figure}

\begin{figure}[tbp]
\centering 
\includegraphics[width=\textwidth]{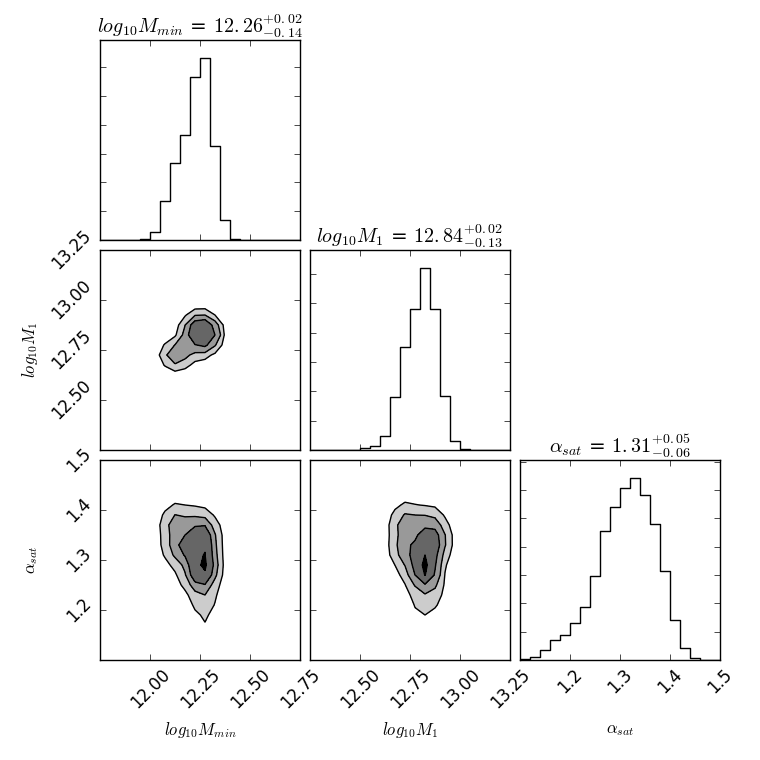}
\caption{\label{fig:ac_corner} Estimated background auto-correlation MCMC free parameters posterior distributions (see Section \ref{sec:theo} and Eq. \ref{eq:hod} for free parameters' explanation). The contours levels correspond to $\sim$12\%, $\sim$39\%, $\sim$68\% and $\sim$86\% of the posterior area.}
\end{figure}

In order to check that both our samples have no special statistical properties that make them extraordinary, we estimate their angular auto-correlation to compare it with previous determinations.

The angular (auto)-correlation function, $w(\theta)$, is a measure of the probability, in excess of the expectation for a Poisson distribution, of finding a galaxy within each of two infinitesimal solid angles 
separated by an angle $\theta$, $P(\theta) = N[1 + w(\theta)]$, where $N$ is mean the surface density of galaxies. We have computed $w(\theta)$ for the background and foreground samples using the \cite{Lan93} estimator
\begin{equation}
w(\theta)={{\rm DD}(\theta)-2{\rm DR}(\theta)+{\rm RR}(\theta)\over {\rm RR}(\theta)},
\end{equation}
where DD$(\theta)$ is the normalized number of unique pairs of real sources with separation $\theta$, DR$(\theta)$ is the normalized number of unique pairs between the real catalogue and a mock sample of sources with random positions, and RR$(\theta)$ is the normalized number of unique pairs in the random source catalogue.

The total common area between both samples was divided in several circular mini-regions with radius of 120 arcmin. In order to maximize the usable areas we allowed a maximum overlap between mini-regions of 30\% and less than 10\% of each mini-region area without sources (to deal with borders and sample irregularities). These constraints provide us a list of 16 usable almost independent mini-regions.

We computed the auto-correlation of GAMA galaxies (\textit{foreground sample}) for each mini-region. By considering each measurement as independent, our measured angular auto-correlation function is simply the average over the mini-regions computed values at each angular scale. The uncertainties are the standard error of the mean, $s/\sqrt{N}$, with $s$ the sample standard deviation and $N=16$ the number of mini-regions. Our estimated angular auto-correlation is in line with previous determinations carried out for the $r$-band magnitudes interval $18<z<19$ using the full SDSS catalogue \cite{Con02,Wan13} (see Fig.~\ref{fig:ac}; \textit{top panel}).

The angular auto-correlation function of our background sample is shown in Fig.~\ref{fig:ac} (\textit{red circles; bottom panel}). 
The computed auto-correlation was limited to angular scales $\gtrsim30$ arcsec to avoid the potential bias caused by the resolution of the instruments (FWHM$\sim18$ and 25 arcsec for the 250 and $350\mu$m bands respectively; \cite{Rig11,Val16}). The signal is clearly detected up to scales $\gtrsim 50$\,arcmin; it is dominated by the 2-halo term on scales above $\simeq 2$\,arcmin and by the 1-halo term on smaller scales. We compare it with the auto-correlation estimated by GN14 (\textit{gray squares}). The updated auto-correlation is measured with better accuracy as expected from the new catalogues and sample selection and it is in good agreement with the previous estimation.

Taking into account the relatively large area of the mini-regions, $\sim 12.6\,\rm{deg}^2$, we have considered the ``integral constraint'' correction negligible. Similar conclusion can be obtained by comparing the foreground auto-correlation with previous works \cite{Con02,Wan13} or the background one with the theoretical expectations.

We take the opportunity to apply the \textit{Markov Chain Monte Carlo} (MCMC) framework, that we set up to analyze the cross-correlation signal (see section \ref{sec:xchm} for a full description), to the analysis of the auto-correlation signal. The halo modeling of the auto-correlation signal is straightforward and commonly used in literature and can be used to check the validity of our pipelines. In the case of the auto-correlation signal, the power spectrum can be expressed as (see \cite{Coo02} for example):
\begin{subequations}
\begin{equation}
P_\mathrm{gal}(k,z) = P^{1h}_\mathrm{gal}(k,z) + P^{2h}_\mathrm{gal}(k,z),
\end{equation}
where
\begin{equation}
P^{1h}_\mathrm{gal}(k,z) = \int{dM \frac{dN}{dM}(z)}\frac{\langle N_\mathrm{gal}(N_\mathrm{gal}-1) | M\rangle}{\overline{n}_\mathrm{gal}^2} \vert u_\mathrm{gal} (k|M,z)\vert^{p},
\end{equation}
\begin{equation}
P^{2h}_\mathrm{gal}(k,z) = P^\mathrm{lin}(k,z) \left[ \int{dM \frac{dN}{dM}(z)\frac{\langle N_\mathrm{gal}|M\rangle}{\overline{n}_\mathrm{gal}} b(M,z)u_\mathrm{gal}(k|M,z)}\right]^2.
\end{equation}
\end{subequations}

Moreover, the auto-correlation simplifies to:
\begin{equation}
\omega(\theta)= \int dz\frac{H_0 E(z)}{c\, \chi^2(z)}\, \left( \frac{dN}{dz}\right)^2 \int_0^\infty{\frac{\ell d\ell}{2\pi}P_\mathrm{gal}(\ell/\chi(z),z)J_0(\ell\theta)}
\end{equation}

As described in section \ref{sec:theo}, we adopt the same halo occupation distribution expressed in eq. \ref{eq:hod}.

In our auto-correlation analysis the free parameters are $M_{min}$, $M_1$ and $\alpha_{sat}$. Notice that, in works as \cite{Xia12}, only two free parameters are analyzed, by considering $M_1=20M_{min}$. For our free parameters we decided to use a non-informative, or uniform/flat, priors: $10<log_{10}(M_{min}/M_\odot) < 13$, $11<log_{10}(M_1/M_\odot) < 14$ and $0.5<\alpha_{sat} < 3$. The MCMC results can be seen in Fig. \ref{fig:ac_corner}. The best fit values are (mean and 68\% confidence interval) : $log_{10}(M_{min}/M_\odot)=12.26^{+0.02}_{-0.14}$, $log_{10}(M_1/M_\odot)=12.84^{+0.02}_{-0.13}$ and $\alpha_{sat}=1.31^{+0.05}_{-0.06}$.

The SPIRE resolution limits the estimation of the auto-correlation function to scales higher than $\sim20"$. For this reason, the 1-halo dominance region is not well sampled causing a small degeneracy between the $M_1$ and $\alpha_{sat}$ parameters.

The current constraints derived from the halo model analysis of the auto-correlation measurement are in good agreement with the results by \cite{Xia12} ($log_{10}(M_{min}/M_\odot)=12.24\pm0.06$ and $\alpha_{sat}=1.81\pm0.04$) and, in particular, with \cite{Coo10} findings ($log_{10}(M_{min}/M_\odot)=12.6^{+0.3}_{-0.6}$, $log_{10}(M_1/M_\odot)=13.1^{+0.3}_{-0.5}$ and $\alpha_{sat}=1.3\pm0.4$). Similar conclusions were obtained by \cite{Mit12,Wil17} using cross-correlation measurements of high redhsift sub-millimetre galaxies and optical surveys to derive more precise angular correlation functions.

Recently, \cite{Che16b} have measured the correlation length of a sample of  $\sim 3000$ sub-millimetre galaxies, with redshifts $z \sim$ 1 -- 5 and star formation rates $\gtrsim 60$ -- 100 $M_\odot /yr$, identified using a new colour selection technique, which combines three optical-near-infrared colors \cite{Che16a}. In Figure \ref{fig:ac} is also plotted the auto-correlation measured by \cite{Che16b} for sources with $2<z<3$ (\textit{green diamonds}). It is remarkable the good agreement of our halo model best fit even at angular scales not accessible with H-ATLAS data.

Therefore, the main conclusion obtained by analyzing the auto-correlation signal of our samples is that they do not show any special statistical properties with respect other samples used for this kind of studies. Moreover, we checked that the MCMC framework and halo modeling are robust and reliable and can be safely applied to the analysis of the measured cross-correlation signal.

\subsection{Angular cross-correlation}\label{sec:xc}

\begin{figure}[tbp]
\centering 
\includegraphics[width=\textwidth]{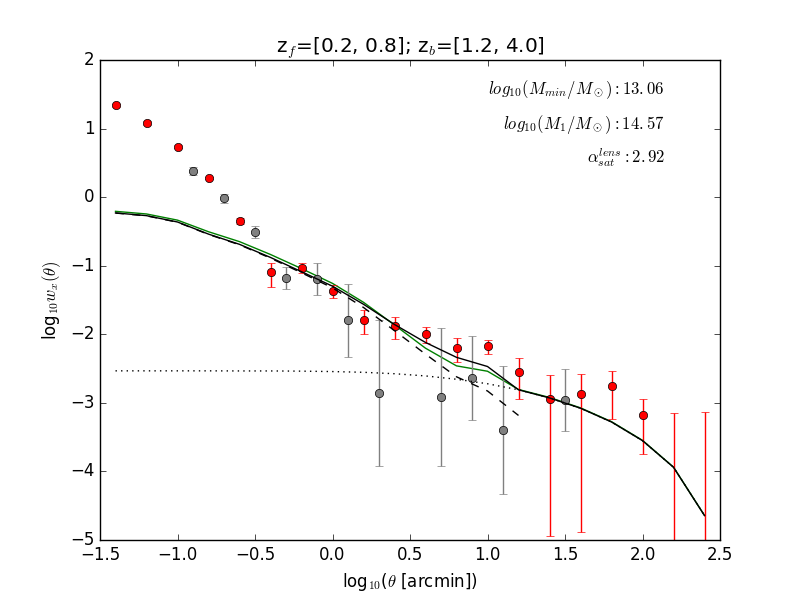}
\caption{\label{fig:xc} Estimated angular cross-correlation between the GAMA (foreground) and the H-ATLAS (background) samples (\textit{red circles}). The gray points correspond to the measurements obtained by GN14. The best halo model fit is shown as black lines (total, \textit{solid}; 1-halo, \textit{dashed} and 2-halo, \textit{dotted}). The green line indicates the best fit when considering an $\alpha_{sat}\sim 1.8$ gaussian prior. Measurements below $\sim 30$ arcsec are not used in the model fit (see text).}
\end{figure}

The cross-correlation function of two source populations is the fractional excess probability, relative to a random distribution, of finding a source of population 1 and a source of population 2, respectively, within infinitesimal solid angles
separated by an angle $\theta$ \cite{Pee80}. We have computed the cross-correlation between our background and foreground samples using a modified version of the \cite{Lan93} estimator \cite{Her01}:
\begin{equation}
w_x(\theta)=\frac{\rm{D}_1\rm{D}_2-\rm{D}_1\rm{R}_2-\rm{D}_2\rm{R}_1+\rm{R}_1\rm{R}_2}{\rm{R}_1\rm{R}_2}
\end{equation}
where $\rm{D}_1\rm{D}_2$, $\rm{D}_1\rm{R}_2$, $\rm{D}_2\rm{R}_1$ and $\rm{R}_1\rm{R}_2$ are the normalized data1-data2, data1-random2, data2-random1 and random1-random2 pair counts for a given separation $\theta$ (see \cite{Bla06} for a discussion of different estimators of $w_x(\theta)$).

We have followed the same procedure as in the auto-correlation case by computing the angular cross-correlation function in the 16 circular mini-regions and estimating the mean values and their associated standard errors. With this procedure we are trying to minimize the sample variance effect. As in the auto-correlation case we have considered the ``integral constraint'' correction negligible due to the relative large are of each mini-region ($\sim 12.6\, \rm{deg}^2$).

The measured angular cross-correlations between the foreground ($0.2<z<0.8$) and the background ($1.2<z<4.0$) samples are shown in Fig.~\ref{fig:xc} (\textit{red circles}). Unlike the auto-correlation case, the small angular scale limit is dictated by the H-ATLAS positional error (the SDSS one is negligibly small compared to it) whose rms value at $250\,\mu$m is $\sim2.4\,$arcsec for $5\,\sigma$ sources \cite{Rig11,Smi11,Val16}. In fact, below $\sim 6"$ we can appreciate the gaussian filtering effect produced by the positional uncertainties: the first two points are departing from a power law indicating a convergence toward a saturation value at smaller angular distances. Our results are in very good agreement with those of GN14 (\textit{gray squares}), although with better accuracy and wider angular scales coverage thanks to the improvements in the catalogue production, sample selection criteria, wider area and tailored analysis pipeline. The mean values and 68\% confidence intervals of the cross-correlation results can be found in Table~\ref{tab:xc_data}.

\begin{table}[tbp]
\centering
\begin{tabular}{|c|r|r|r|r|r|}
\hline
$log_{10}(\theta)$ & \multicolumn{5}{|c|}{$log_{10}(w_x)$} \\
\cline{2-6}
$\left[\rm{arcmin}\right] $ & 0.2<z<0.8 & 0.1<z<0.2 & 0.2<z<0.3 & 0.3<z<0.5 & 0.5<z<0.8 \\
\hline
-1.4 & $ 1.34$\footnotesize$^{+0.02}_{-0.02}$ & $ 1.23$\footnotesize$^{+0.04}_{-0.04}$ & $ 1.21$\footnotesize$^{+0.03}_{-0.03}$ & $ 1.39$\footnotesize$^{+0.03}_{-0.03}$ & $ 1.64$\footnotesize$^{+0.05}_{-0.05}$ \\
-1.2 & $ 1.09$\footnotesize$^{+0.03}_{-0.03}$ & $ 0.96$\footnotesize$^{+0.02}_{-0.03}$ & $ 0.97$\footnotesize$^{+0.03}_{-0.03}$ & $ 1.14$\footnotesize$^{+0.03}_{-0.04}$ & $ 1.32$\footnotesize$^{+0.04}_{-0.04}$ \\
-1.0 & $ 0.74$\footnotesize$^{+0.02}_{-0.02}$ & $ 0.57$\footnotesize$^{+0.03}_{-0.03}$ & $ 0.68$\footnotesize$^{+0.03}_{-0.03}$ & $ 0.77$\footnotesize$^{+0.03}_{-0.03}$ & $ 0.79$\footnotesize$^{+0.04}_{-0.04}$ \\
-0.8 & $ 0.28$\footnotesize$^{+0.02}_{-0.02}$ & $ 0.28$\footnotesize$^{+0.03}_{-0.03}$ & $ 0.29$\footnotesize$^{+0.03}_{-0.03}$ & $ 0.25$\footnotesize$^{+0.03}_{-0.03}$ & $ 0.37$\footnotesize$^{+0.05}_{-0.06}$ \\
-0.6 & $-0.35$\footnotesize$^{+0.04}_{-0.05}$ & $-0.40$\footnotesize$^{+0.04}_{-0.04}$ & $-0.41$\footnotesize$^{+0.05}_{-0.06}$ & $-0.32$\footnotesize$^{+0.07}_{-0.08}$ & $-0.21$\footnotesize$^{+0.08}_{-0.10}$ \\
-0.4 & $-1.09$\footnotesize$^{+0.14}_{-0.21}$ & <-1.55 & $-1.06$\footnotesize$^{+0.13}_{-0.19}$ & $-1.13$\footnotesize$^{+0.23}_{-0.50}$ & $-0.86$\footnotesize$^{+0.18}_{-0.31}$ \\
-0.2 & $-1.03$\footnotesize$^{+0.06}_{-0.07}$ & $-1.13$\footnotesize$^{+0.12}_{-0.17}$ & $-1.27$\footnotesize$^{+0.12}_{-0.17}$ & $-0.95$\footnotesize$^{+0.10}_{-0.12}$ & $-0.71$\footnotesize$^{+0.10}_{-0.14}$ \\
-0.0 & $-1.37$\footnotesize$^{+0.09}_{-0.11}$ & $-1.62$\footnotesize$^{+0.19}_{-0.34}$ & $-1.63$\footnotesize$^{+0.21}_{-0.43}$ & $-1.31$\footnotesize$^{+0.10}_{-0.13}$ & $-1.07$\footnotesize$^{+0.12}_{-0.16}$ \\
 0.2 & $-1.79$\footnotesize$^{+0.14}_{-0.20}$ & $-1.58$\footnotesize$^{+0.16}_{-0.26}$ & $-1.94$\footnotesize$^{+0.23}_{-0.51}$ & $-1.87$\footnotesize$^{+0.16}_{-0.26}$ & $-1.22$\footnotesize$^{+0.12}_{-0.16}$ \\
 0.4 & $-1.88$\footnotesize$^{+0.13}_{-0.18}$ & $-1.83$\footnotesize$^{+0.21}_{-0.44}$ & $-2.12$\footnotesize$^{+0.28}_{-1.00}$ & $-1.82$\footnotesize$^{+0.11}_{-0.14}$ & $-1.56$\footnotesize$^{+0.15}_{-0.23}$ \\
 0.6 & $-2.00$\footnotesize$^{+0.10}_{-0.13}$ & $-2.09$\footnotesize$^{+0.19}_{-0.34}$ & $-2.00$\footnotesize$^{+0.13}_{-0.18}$ & $-2.12$\footnotesize$^{+0.27}_{-0.92}$ & $-1.78$\footnotesize$^{+0.13}_{-0.19}$ \\
 0.8 & $-2.20$\footnotesize$^{+0.14}_{-0.21}$ & $-2.26$\footnotesize$^{+0.17}_{-0.30}$ & $-2.20$\footnotesize$^{+0.22}_{-0.48}$ & $-2.18$\footnotesize$^{+0.17}_{-0.27}$ & $-2.09$\footnotesize$^{+0.24}_{-0.58}$ \\
 1.0 & $-2.17$\footnotesize$^{+0.09}_{-0.12}$ & $-3.39$\footnotesize$^{+0.99}_{-2.00}$ & $-2.22$\footnotesize$^{+0.13}_{-0.19}$ & $-2.25$\footnotesize$^{+0.18}_{-0.32}$ & $-1.87$\footnotesize$^{+0.13}_{-0.20}$ \\
 1.2 & $-2.55$\footnotesize$^{+0.21}_{-0.41}$ & <-2.45 & $-2.94$\footnotesize$^{+0.53}_{-2.00}$ & $-2.38$\footnotesize$^{+0.16}_{-0.26}$ & $-2.21$\footnotesize$^{+0.15}_{-0.24}$ \\
 1.4 & $-2.94$\footnotesize$^{+0.35}_{-2.00}$ & <-2.61 & $-3.28$\footnotesize$^{+0.68}_{-2.00}$ & $-2.86$\footnotesize$^{+0.36}_{-2.00}$ & $-2.42$\footnotesize$^{+0.23}_{-0.51}$ \\
 1.6 & $-2.88$\footnotesize$^{+0.30}_{-2.00}$ & <-2.69 & $-3.11$\footnotesize$^{+0.53}_{-2.00}$ & $-2.79$\footnotesize$^{+0.26}_{-0.78}$ & $-2.64$\footnotesize$^{+0.32}_{-2.00}$ \\
 1.8 & $-2.75$\footnotesize$^{+0.22}_{-0.48}$ & $-2.65$\footnotesize$^{+0.22}_{-0.45}$ & $-2.75$\footnotesize$^{+0.29}_{-1.21}$ & $-2.84$\footnotesize$^{+0.24}_{-0.58}$ & $-2.58$\footnotesize$^{+0.23}_{-0.51}$ \\
 2.0 & $-3.18$\footnotesize$^{+0.24}_{-0.57}$ & $-2.57$\footnotesize$^{+0.19}_{-0.33}$ & $-3.20$\footnotesize$^{+0.31}_{-2.00}$ & $-3.66$\footnotesize$^{+0.70}_{-2.00}$ & $-2.76$\footnotesize$^{+0.16}_{-0.24}$ \\
 2.2 & <-3.15 & <-3.16 & <-2.98 & <-3.17 & <-2.91 \\
 2.4 & <-3.14 & <-2.88 & <-3.06 & <-2.94 & <-2.81 \\
\hline
\end{tabular}
\caption{\label{tab:xc_data} Measured angular cross-correlation signal (mean and 68\% confidence intervals) for the different redshift selected samples.}
\end{table}

We refer the reader to \cite{Lap11,GN12,GN14} for a detailed discussion on the low expected level of cross-contamination (observational constrains, catastrophic photo-z failures, physical Spectral Energy Distribution analysis, etc) between the foreground and background samples. In this work we just confirmed their findings by performing a simple but powerful test: to estimate the angular cross-correlation modifying the lower redshift limits of the background sample. In case of a non negligible cross-contamination between both samples, a modification of the lower redshift limit of the background sample would imply a strong variation in the measured cross-correlation signal. Using $1<z<4$ and $1.5<z<4$ to select the background sample did not introduce any noticeable difference respect our default measured cross-correlation (see next section for an additional test). Although allowing background sources with z>1 decrease a little the measurements uncertainties, we preferred to maintain the more conservative, and almost equally accurate, default redshift range ($1.2<z<4$) for the background sample.

Before any attempt of a halo modeling (see section \ref{sec:xchm}), we can already identify three different regimes in the estimated cross-correlation signal. For $\theta> 10'$ (2 -- 3 Mpc at $z\sim0.3$) we observe the correlation decline expected from the 2-halo term (the background galaxies embedded in dark matter halos suffer weak lensing amplifications produced by those foreground dark matter halos that are positionally correlated with them). Therefore, at smaller angular scales we expect the dominance of the 1-halo term (background galaxies that suffer a weak lensing amplification by foreground sub-halos inside the correlated halos described in the 2-halo term). However, below $\theta\lesssim 30''$ ($\lesssim 135$ kpc at $z\sim0.3$) we find a clear change in the slope of the signal, not easily explained within the traditional 1-halo term. We interpret this excess as the effect caused by the strong lensing produced by the more massive foreground galaxies. We will discuss this point in more detail in section \ref{sec:strlen}.

\subsection{Tomographic angular cross-correlation}

\begin{figure}[tbp]
\centering 
\includegraphics[width=\textwidth]{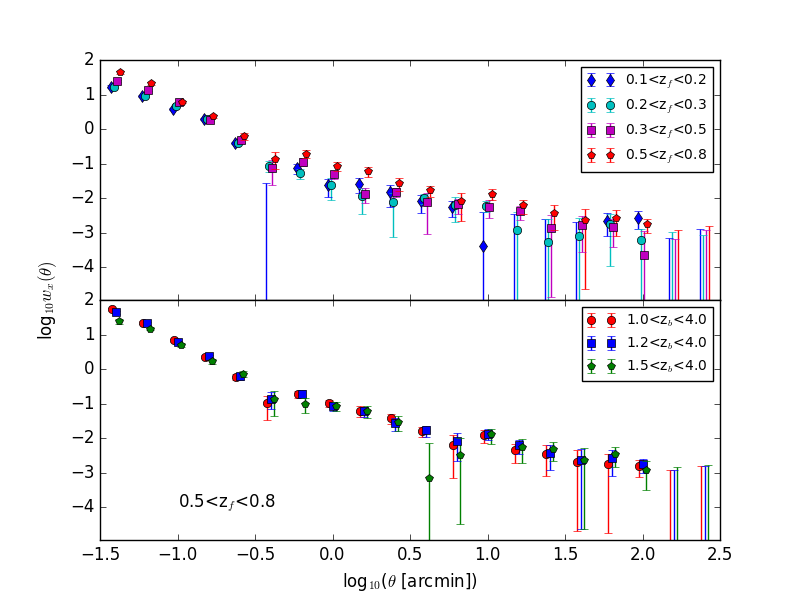}
\caption{\label{fig:xc_z} \textit{Top panel}: Tomographic measurements of the angular cross-correlation for different foreground redshift bins. \textit{Bottom panel}: Study of the variation of the measured cross-correlation signal with the lower redshift limit of the background sample for the foreground bin $0.5<z<0.8$, the more sensitive one to any cross-contamination effect. We introduced a small x-axis offset in each sample for visibility purposes.}
\end{figure}

As stated before, GN14 reported a highly significant correlation between the spatial distribution of H-ATLAS galaxies with estimated redshift $z>1.5$ and that of SDSS/GAMA galaxies at $0.2 < z < 0.6$. The much higher significance compared to those reported so far is a result of the extreme steepness of the sub-mm source counts. The fact that such high significance was obtained using less than 10\% the number of sources typically involved in the optical results opens the possibility of tomographic analyses.

Nowadays, the concept of a tomographic analysis represents the idea of sectioning an image or sample in different slices to perform a more detailed study. In our case, we split the foreground sample in three different redshift intervals ($0.2<z<0.3$, $0.3<z<0.5$ and $0.5<z<0.8$ ), and we consider also an additional one at lower redshift ($0.1<z<0.2$). The redshift ranges were selected in order to maintain an acceptable number of foreground galaxies. Taking into account that all the redshifts were obtained from spectroscopic surveys, we simply neglect any possible mismatching between redshift slices.

We applied the same procedure used to estimate the angular cross-correlation signal for the default foreground sample. The measured angular cross-correlations  signals are written in Table \ref{tab:xc_data}. As can be seen in Fig. \ref{fig:xc_z} (\textit{top panel}), we have a clear detection of the measured cross-correlation signal in all four redshift bins. Splitting the sample in three different slices increases the uncertainty of the measurements, as expected. By comparing the four measured signals we can notice that there is an increase of power with redshift.

Due to the lower number of foreground galaxies and it proximity to the background lower redshift limits, the foreground redshift bin $0.5 <z<0.8$ is the one that can be most affected by any possible cross-contamination issue. Figure \ref{fig:xc_z} (\textit{bottom panel}) shows the robustness of the measurements in this redshift bin against a modification of the lower redshift limit of the background sample (as in the previous subsection). Taking into account the lower number of foreground galaxies, the effect of a small number of catastrophic photo-z failures should be noticeable in this test. Therefore we can confirm again that the potential cross-contamination between both samples is negligible.

\section{Discussion}

This section is dedicated to describe in detail the halo modeling applied to all the measured cross-correlation signals and to discuss the astrophysical constraints that can be derived.

\begin{figure}[tbp]
\centering 
\includegraphics[width=\textwidth]{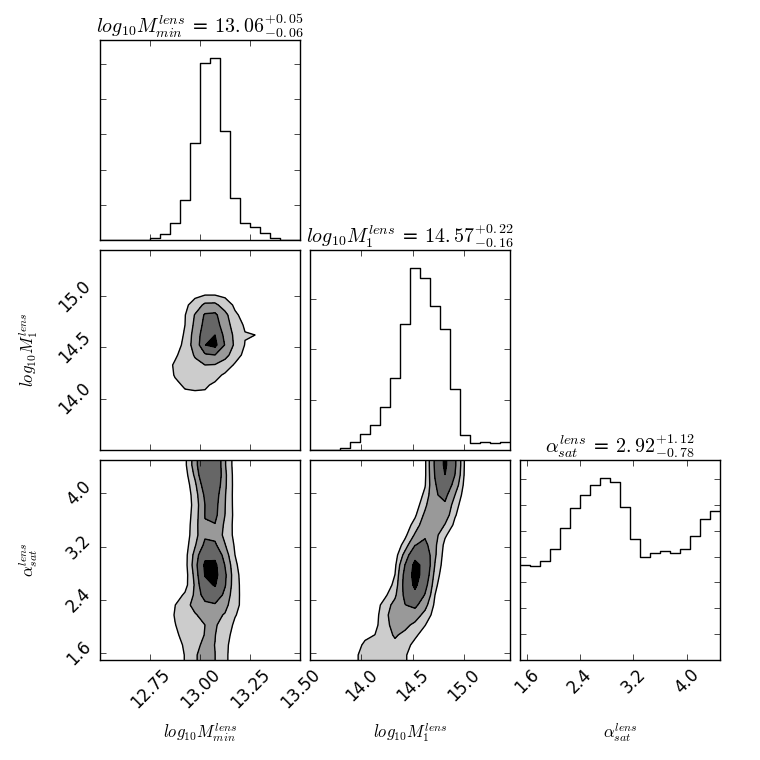}
\caption{\label{fig:corner} Estimated cross-correlation free parameters posterior distributions for the default foreground sample (lenses; $0.2<z_f<0.8$). The contours levels correspond to $\sim$12\%, $\sim$39\%, $\sim$68\% and $\sim$86\% of the posterior area.}
\end{figure}

\subsection{Weak lensing regime}\label{sec:xchm}
Section \ref{sec:theo} describes in detail the assumptions and halo model formalism that we will apply to our measured cross-correlation signals. We decided to remain with the most simplistic descriptions of important quantities (the HOD in eq. \ref{eq:hod}, for example) in order to have a lower number of free parameters and a simpler interpretation of our findings. 

Following \cite{Bia15,Bia16} we adopted $\beta = 3$ as our fiducial value. Therefore, as briefly described in section \ref{sec:ac}, we are left just with three free parameters: $M_{min}^{lens}$, $M_1^{lens}$ and $\alpha_{sat}^{lens}$ (we added the superscript to distinguish them from the auto-correlation ones due to their different physical interpretations as described below).  We constrain them by comparing our theoretical model with the measured signal by means of a MCMC framework. Since we consider the measurements below $\sim 30$ arcsec produced mainly by the strong lensing effect, the adopted model can not be a good description of the data in this regime. Therefore, we do not use them when performing the model fits.

In this work we make use of the open source \texttt{PyMC}\footnote{https://pymc-devs.github.io/pymc/} software package. \texttt{PyMC} is a python module that implements Bayesian statistical models and fitting algorithms, including Markov chain Monte Carlo. We decided to use a non-informative, or uniform/flat, priors for our free parameters: $11<log_{10}(M_{min}^{lens}/M_\odot) < 15$, $12<log_{10}(M_1^{lens}/M_\odot) < 15.5$ and $1.5<\alpha_{sat}^{lens} < 4.5$. For each signal analysis we generated at least 10000 posterior samples to ensure good statistical sampling after convergence.

The black lines (total, \textit{solid}; 1-halo, \textit{dashed}; 2-halo, \textit{dotted}) in Fig. \ref{fig:xc} represent the halo model best fit estimated using the MCMC approach. The best fit posterior distributions of the free parameters for the main foreground lensing sample can be seen in Fig. \ref{fig:corner}. The best fit values are (mean and 68\% confidence intervals): $log_{10}(M_{min}^{lens}/M_\odot)=13.06^{+0.05}_{-0.06}$ and $log_{10}(M_1^{lens}/M_\odot)=14.57^{+0.22}_{-0.16}$ (see also Table \ref{tab:bfv}). The halo model best fit parameters produce a very good fit to the data (at least above $\sim 30"$). Moreover, their posterior distributions show the constraints achievable by analyzing the cross-correlation signal (see Fig. \ref{fig:corner}).
For the particular case of the $\alpha_{sat}^{lens}$ parameter we obtain a best fit value of $\alpha_{sat}^{lens}=2.92^{+1.12}_{-0.78}$, although it is clear from Fig. \ref{fig:corner} that it is almost unconstrained and completely dependent on the flat prior (see detailed discussion below).

\begin{figure}[tbp]
\centering 
\includegraphics[width=\textwidth]{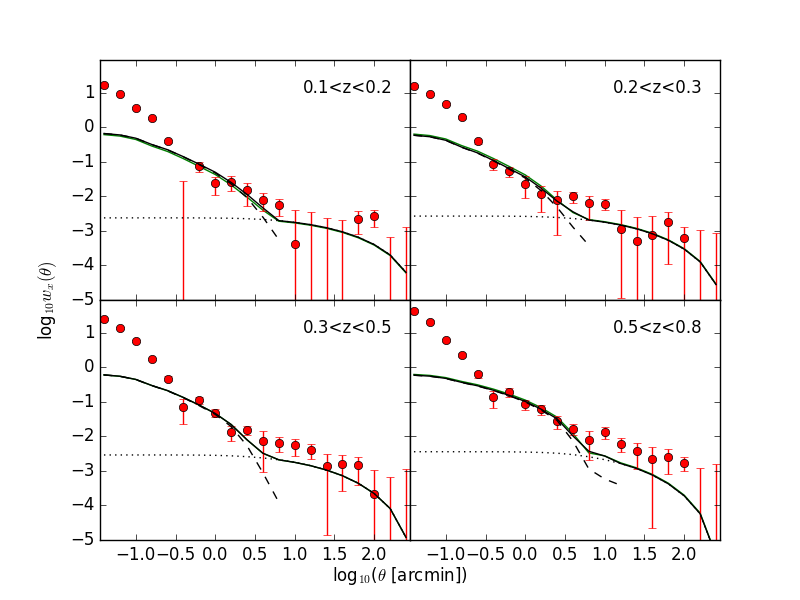}
\caption{\label{fig:xcz} Angular cross-correlation between the H-ATLAS background sample ($1.2 <z < 4.0$) and the GAMA foreground one, divided in three redshift bins ($0.2 < z < 0.3$, \textit{top right}; $0.3 < z < 0.5$, \textit{bottom left}; $0.5 < z < 0.8$, \textit{bottom right}) plus and additional sub-sample at lower redshift ($0.1 < z < 0.2$, \textit{top left}). The best halo model fit for each redshift bin is shown as black lines (total, \textit{solid}; 1-halo, \textit{dashed} and 2-halo, \textit{dotted}) in each panel. The green lines indicate the best-fits when considering an $\alpha_{sat}^{lens}\sim 1.8$ gaussian prior. Measurements below $\sim 30$ arcsec are not used in the model fits (see text).}
\end{figure}

Taking into account that this signal is produced \textit{only} by the gravitational lensing effect, the $M_{min}^{lens}$, represents the minimum mass of a dark matter halo acting as a deflector that is able to produce an statistical measurable weak lensing amplification on the background sources. These halos are pinpointed by massive galaxies in their centers from our foreground sample. At certain mass, $M_1^{lens}$, these halos contain satellite sub-halos massive enough that, in turn, became also deflectors producing a statistical measurable weak lensing effect. Finally, the $\alpha_{sat}^{lens}$ parameter is simply the rate of the increase of satellite \textit{deflectors} with mass above $M_1^{lens}$ and, as seen in Fig \ref{fig:corner}, it not well constrained with our current data.

These results are agreement with the main conclusions obtained by GN14. Based on  realistic simulations of clustered sub-mm galaxies amplified by foreground structures, they were able to confirm that the cross-correlation can be explained by weak gravitational lensing ($ \mu< 2$). The simulations also showed that the signal can be reproduced if SDSS/GAMA galaxies act as signposts of galaxy groups/clusters with halo masses in the range $10^{13.2}-10^{14.5} M_\odot$. There is a particularly remarkable agreement between the minimum mass derived from simulations and the current estimated value obtained from the halo model fitting to the main sample measured cross-correlation signal.

However, these results are higher than the traditional values obtained from \textit{normal} galaxies studied in the optical band (galaxies with stellar mass, $log_{10}(M_\star/M_\odot)< 11$ \cite{Hat16} or luminous red galaxies, LRG, with $M_r>-21$ \cite{Zhe09}): $log_{10}(M_{min}/M_\odot)\sim12.5$, $log_{10}(M_1/M_\odot)\sim 13.8$ and $\alpha_{sat}\sim 1.0$. In addition, it was also found an almost constant relationship between both HOD masses: $M_1/M_{min}=10-30$. Our results are more in agreement when they are compared with the values derived from massive galaxies, $log_{10}(M_\star/M_\odot)> 11$ \cite{Mat11} or the brightest LRGs, $M_r<-21$ \cite{Zhe09}. In the first case, the $M_{min}$ value is significantly larger for more massive galaxies ($\sim 10^{13 - 14.5} M_\odot$), which is consistent with the general agreement that the capability of hosting massive galaxies depends strongly on halo mass. Moreover, the $M_1$ mass is extremely large ($>10^{14.5} M_\odot$), which means that only the most massive halos could host a galaxy with the stellar mass exceeding $10^{11} M_\odot$ as a satellite. Similarly,  most of the LRGs in the brightest luminosity bin, $M_r<-21$, are central galaxies in halos of mass $\sim 10^{13.3 - 14.3} M_\odot$, and a small fraction, $\sim∼7\%$, of them are satellites in more massive halos. These observational constraints reinforce the lensing influence in our results, i.e. although our foreground sample is not particularly special from the mass or luminosity point of view, the cross-correlation signal is produced only by the lensing effect caused by the more massive galaxies among it, mostly central galaxies. The other galaxies can be considered almost non existent.

The higher $\alpha_{sat}^{lens}$ value deserve a special discussion. In studies of bright LRG, the inferred high mass slopes of their occupation functions tend to be substantially larger than unity ($\alpha_{sat} >1.5$, see \cite{Zeh11} and reference therein) as determined from observational inferences and theoretical predictions for low luminosity samples. They also verified that the predicted number of LRGs in high mass halos appears approximately consistent with the observational estimates from massive clusters. Similarly, in the case of $M_\star > 10^{11} M_\odot$ galaxies between $0.2<z<1$, \cite{Mat11} found consistently $\alpha_{sat}=1.5-2.0$ values in agreement with the LRGs ones. In our case, we find a $\alpha_{sat}^{lens}$ value even higher but it can be considered almost unconstrained and prior dependent, taking into account the posterior distribution shown in Fig. \ref{fig:corner}. The main reason is that the effect of the $\alpha_{sat}^{lens}$ parameter is most obvious around $3-10$ arcmin, the scales between the 1-halo and 2-halo transition region, while the masses $M_{min}^{lens}$ and $M_1^{lens}$ are constrained by the largest and smallest scales respectively. It is clear in Fig. \ref{fig:xc} that the observed correlation around $\sim10$ arcmin is stronger than expected from the 2-halo term, indicating the necessity of higher $\alpha_{sat}^{lens}$ values.

In order to evaluate the robustness of our results with respect the particular $\alpha_{sat}^{lens}$ value we performed an additional test: we estimated again the HOD masses but using a gaussian prior for $\alpha_{sat}^{lens}$ with a mean value of $1.8$ and a $0.3$ dispersion, following \cite{Mat11} results. The new estimated masses are compared with the previous ones in Table \ref{tab:bfv} and the best fit is represented as a green line in Fig. \ref{fig:xc}. As anticipated in the previous paragraph, the estimated HOD masses are almost independent of the $\alpha_{sat}^{lens}$ parameter. Therefore, although not relevant for the study of the other general HOD properties, the smooth transition between the 1-halo and 2-halo terms observed in the clustering of massive galaxies could be indicative of the excessive simplicity of the HOD framework, as already pointed out in several contexts (e.g. \cite{Puj17}).

\begin{figure}[tbp]
\centering 
\includegraphics[width=\textwidth]{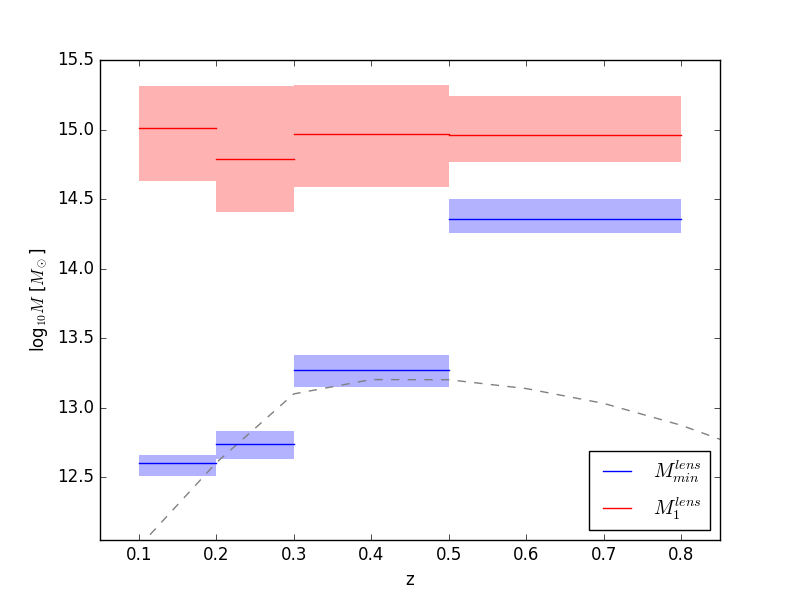}
\caption{\label{fig:zevo} Redshift evolution of the HOD parameters derived from the four redshift bins foreground sub-samples. The theoretical estimation of the evolution with redshift of the statistical deflector mass, following \cite{Lap12} formalism, is shown as gray dashed line in the top panel (see text for more details).}
\end{figure}

When we apply the same halo model fit methodology to the angular cross-correlation signals measured from the four foreground redshift bins (see Fig. \ref{fig:xcz}), we obtain the best fit values shown in Table \ref{tab:bfv}. They are also summarized in Fig. \ref{fig:zevo}. As expected, the results from the default foreground sample ($0.2<z<0.8$) are close to a median results when compared with the redshift sub-samples ones. The parameters uncertainties for each redshift bin are comparable, confirming that the number of foreground galaxies in each redshift bin is enough for our purposes.

The tomographic results show a clear evolution with redshift of the minimum mass needed to produce a measurable weak lensing effect (from $log_{10} \left( M_{min}^{lens}/M_\odot\right) = 12.61^{+0.06}_{-0.09}$ for $0.1<z<0.2$ to $14.36^{+0.14}_{-0.10}$ for $0.5<z<0.8$). Although a similar trend was found by \cite{Mat11} studying the clustering properties of massive galaxies, $M_\star > 10^{11} M_\odot$, at $z>0.2$, we think that in our case this redshift evolution is not an intrinsic property of the sample but a direct effect of the dependence of the lensing probability with redshift. The lensing probability increases steeply with redshift, with a maximum around $z\sim 0.5$ for a typical background source at $z\sim2$ (see \cite{Lap12}, for example). In other words, in order to produce a statistical gravitational effect almost independent of redshift, similar to the measured cross-correlations functions, we can reduce the number of deflectors required, thanks to the higher lensing probability. A lower number of deflectors implies a higher halo mass, as indicated by a typical halo mass function.

Using \cite{Lap12} formalism we derive the theoretical evolution expected for the deflector halo mass if the lensed galaxy is situated at $z=2$. The inner integral of the lensing optical depth (eq. 21 in \cite{Lap12}) corresponds to the lens redshift distribution for a fixed minimum deflector mass. We consider only the weak lensing case with $\mu<2$ (i.e. the lens probability for $\mu>1$ minus the lens probability for $\mu>2$). Based on our inferred masses using the default sample, we then estimate the lens redshift distribution for $log_{10} \left( M_{min}^{lens}/M_\odot\right) = 13.1$. From this distribution we obtain the lens probability associated with this minimum deflector mass for $z=0.3$, the approximate mean redshift of the default sample.
Taking into account the similar strength of the measured signals in the different redshift bins, we consider a constant lens probability with redshift. Therefore,  the deflector mass evolution with redshift, shown as a gray dashed line in the top panel of Fig. \ref{fig:zevo}, is obtained estimating the minimum mass that produce the same lens probability when varying the redshift: in order to compensate the lower (higher) lens probability at redshift $z<0.3$ ($z>0.3$), we need to increase (decrease) the number of deflectors implying lower (higher) minimum deflector masses.
It provides a very good explanation of the estimated minimum deflector masses until redshift $z\lesssim0.5$. At higher redshifts, $z>0.5$, the results indicate that most of the signal is caused by massive overdensities with more than one deflector. In fact, the \cite{Lap12} formalism assumes only a central deflector per halo, thus it estimates a redshift evolution for $z>0.5$ that requires a lower deflector mass, i.e. higher number of deflectors to compensate the lower lensing probability, at variance with our findings.

On the contrary, both $M_1^{lens}$ and the steepness of the satellites number, $\alpha_{sat}^{lens}$, remains almost constant with redhsift. This implies that, \textit{independently of redshift}, cluster halos of $log_{10} \left( M_1^{lens} /M_\odot\right)\simeq 15$ have a central deflector and, at least, one sub-halo satellite acting also as a deflector.  Above this mass, it seems that the increase in number of satellites acting as deflectors does not evolve with redshift ($\alpha_{sat}^{lens}\simeq 2.8$), although we have to take into account that this parameter is almost unconstrained by our analysis. In fact, using a gaussian prior of $\alpha_{sat}^{lens}\sim 1.8$, as described before, we find almost no significance difference in the recovered HOD masses for the four different redshift sub-samples (see Table \ref{tab:bfv} and best-fits as green lines in each panel of Fig. \ref{fig:xcz}).

However, as explained before, there is an observed relationship between both HOD masses almost constant with redshift, $M_1/M_{min} \sim 10-30$. Below $z\sim 0.3$ our results indicates a much higher ratio. The reason is again the fact that we are studying the HOD properties through the weak lensing effect. For $z<0.3$ the lens probability decrease very fast and, in order to compensate the lower probability, we need to increase the number of deflectors by going to the more abundant less massive galaxies. The same effect can not be done with the satellites due to their small fraction, $\sim 10\%$. In other words, to maintain an statistically observable magnification bias contribution from the satellites their mass can not be much lower than $M_\star \sim 10^{11} M_\odot$ and therefore this implies $M_1^{lens}$ values above $10^{14.5} M_\odot$.

Finally, there is a discrepancy in the highest redshift sub-sample: the cross-correlation power is stronger than the best-fit model for the 2-halo term. The reason for this discrepancy is already well known (see for example \cite{Mat11} and references therein): the observed cross-correlation function at high redshift is so strong that very massive halos are required to reproduce it, while the predicted number of such massive halos, the high mass behavior of the halo mass function used in our theoretical framework, is very small compared to the observed numbers of massive galaxies. This discrepancy could indicate again that our current HOD model is too simplistic (e.g. other halo characteristics such as the mass accretion rate could affect the galaxy formation and alter the observed properties of galaxies within the halo) or that the evolution of halo mass function and/or bias function is not well understood (earlier emergence of massive halos than predicted in the current halo models) or simply be a dependence with cosmology (e.g. the normalization $\sigma_8$ of the matter fluctuation power spectrum).

\begin{table}[tbp]
\centering
\begin{tabular}{|c|r|r|r|r|r|r|}
\hline
 redshift & \multicolumn{2}{|c|}{$log_{10} \left( M_{min}^{lens}/M_\odot\right)$} & \multicolumn{2}{|c|}{$log_{10} \left( M_1^{lens}/M_\odot\right)$} & \multicolumn{2}{|c|}{$\alpha_{sat}^{lens}$} \\
 \cline{2-7}
 range & \small flat prior & \small$\alpha_{sat}^{lens}$ prior & \small flat prior & \small$\alpha_{sat}^{lens}$ prior & \small flat prior & \small$\alpha_{sat}^{lens}$ prior \\
 \hline
 0.2 -- 0.8 & $13.06$\footnotesize$^{+0.05}_{-0.06}$ & $13.13$\footnotesize$^{+0.10}_{-0.10}$ & $14.57$\footnotesize$^{+0.22}_{-0.16}$ & $14.49$\footnotesize$^{+0.66}_{-0.20}$ & $2.92$\footnotesize$^{+1.12}_{-0.78}$ & $1.92$\footnotesize$^{+0.31}_{-0.30}$\\
 0.1 -- 0.2 & $12.61$\footnotesize$^{+0.06}_{-0.09}$ & $12.59$\footnotesize$^{+0.10}_{-0.09}$ & $15.01$\footnotesize$^{+0.30}_{-0.38}$ & $15.11$\footnotesize$^{+0.59}_{-0.60}$ & $2.71$\footnotesize$^{+1.09}_{-0.87}$ & $1.80$\footnotesize$^{+0.31}_{-0.33}$\\
 0.2 -- 0.3 & $12.74$\footnotesize$^{+0.09}_{-0.11}$ & $12.80$\footnotesize$^{+0.12}_{-0.16}$ & $14.79$\footnotesize$^{+0.52}_{-0.38}$ & $15.04$\footnotesize$^{+0.70}_{-0.72}$ & $2.78$\footnotesize$^{+1.22}_{-0.83}$ & $1.81$\footnotesize$^{+0.33}_{-0.31}$\\
 0.3 -- 0.5 & $13.27$\footnotesize$^{+0.11}_{-0.12}$ & $13.27$\footnotesize$^{+0.10}_{-0.13}$ & $14.97$\footnotesize$^{+0.35}_{-0.38}$ & $14.98$\footnotesize$^{+0.68}_{-0.64}$ & $2.89$\footnotesize$^{+1.15}_{-0.89}$ & $1.82$\footnotesize$^{+0.30}_{-0.30}$\\
 0.5 -- 0.8 & $14.36$\footnotesize$^{+0.14}_{-0.10}$ & $14.42$\footnotesize$^{+0.11}_{-0.15}$ & $14.96$\footnotesize$^{+0.28}_{-0.19}$ & $15.09$\footnotesize$^{+0.44}_{-0.40}$ & $3.18$\footnotesize$^{+0.98}_{-1.07}$ & $1.80$\footnotesize$^{+0.33}_{-0.33}$\\
\hline
\end{tabular}
\caption{\label{tab:bfv} Best fit values (mean and 68\% confidence intervals) derived from the halo modelling fit to the measured angular cross-correlation signal for the different redshift selected samples. The derived values for $\alpha_{sat}^{lens}$ are dominated in both cases by the imposed priors.}
\end{table}

\subsection{Strong lensing regime}\label{sec:strlen}

\begin{figure}[tbp]
\centering 
\includegraphics[width=\textwidth]{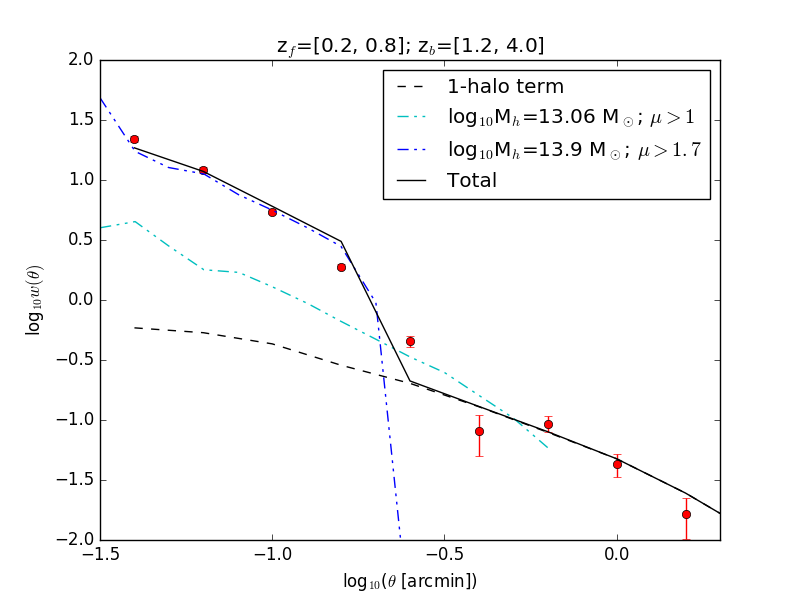}
\caption{\label{fig:strlen} Toy model simulation results to confirm the strong lensing regime  interpretation. The measured angular cross-correlation using the default foreground sample (\textit{red circles}) is fitted with the 1-halo term under the weak lensing approximation (\textit{black dashed line}) plus a contribution of the strong lensing effect, $\mu>1.7$, produced by massive halos, $log_{10}(M_{halo}/M_\odot)=13.9 $, estimated using the simulations described in the text (\textit{blue dot-dot-dashed line}). The simulations results, considering all amplifications, for the derived $M_{min}$ is shown as well (\textit{cyan dot-dot-dashed line})}
\end{figure}

The theoretical formula for the angular cross-correlation between a foreground and background galaxy surveys assumes a weak lensing regime and, therefore, can only be applied to explain the cross-correlation signal produced by weak amplifications (typically much lower than 2). In our case, around $30"$ ($\sim135$ kpc at $z\sim0.3$), we observe a clear steepening that indicates a transition between the weak and strong lensing regimes (see Fig. \ref{fig:xc} and \ref{fig:xcz}). As a consequence, the HOD modelisation provides a good fit only at angular scales greater than $30"$.

Contrary to the auto-correlation case, the 1-halo and 2-halo terms are more related to each other in the cross-correlation halo modeling. As we seen in section \ref{sec:xchm}, this fact put tight constraints to the parameter values. However, there should be no problem to obtain a good enough fit for a 1-halo term with the steepness shown below $30"$. It is the change of slope around such angular scales that makes difficult the halo model fitting under the weak lensing approximation. Different halo density profiles (NFW, singular isothermal sphere or SISSA \cite{Lap12}) can produce noticeable variation at small scales, but for $>30"$ they all provide a single slope very similar to the one derived with our halo model assumptions. Therefore, it can not be a solution for this issue.

On the other hand, taking into account that $\sim 30"$ corresponds to $\sim 135$ kpc at $z\sim0.3$, we are dealing with angular scales comparable with the size of massive galaxies. In fact, as shown by GN14 (see their Fig. 11), it is common for massive galaxies, with halo mass of $log_{10}(M/M_\odot)> 14$, to produce strong lensing amplifications already at angular scales, 20--$30"$, bigger than usual. Therefore, although it is still possible to consider the possibility of satellites at such small physical scales (mainly due to projection effects), it is more natural to interpret this change of slope as the strong lensing effect: faint background sources that by chance appears at such small angular distance from such massive galaxies received an amplification that make them detectable by \textit{Herschel} instruments and increase enormously the probability to find a foreground-background galaxy pairs so close.

After these considerations, we produce simplistic, but effective, simulations in order to further confirm our strong lensing regime interpretation. In the center of a circular sky patch of 1' radius, we situated a foreground dark matter halo of a fixed mass with a NFW mass density profile at $z=0.4$.  Following the source number count model for un-lensed proto-spheroidal galaxies (the type of galaxies that compose our background sample, \cite{Lap11,Cai13,GN12,GN14}) we generate randomly distributed background galaxies with flux densities above 0.1 mJy. Notice that their auto-correlation is completely negligible for our purpose, even more taking into account their relatively low density at such flux densities. For simplicity, we assume all of the background sources at $z=2.5$, considering that the angular distance --- the one relevant for gravitational lensing estimations--- is almost constant above $z>1$. Basing on the \cite{Lap12} gravitational lensing estimation methodology, we calculate the amplification produced by the foreground lens in each of the background sources. Then, we consider as "detected sources" only those background ones with $S>30$ mJy and we count the number of sources (before and after taking into account the gravitational lensing amplification) at a certain radial distance from the center. Due to the low density of background sources in each single simulation, we repeat the same process $10^5$ times, stacking the results (we assume that all the foreground lenses are exactly equal), in order to increase the statistics. Finally, we estimate the cross-correlation signal as $w_x(\theta)= LL(\theta)/UU(\theta)-1$, with LL the number of lensed background sources above the flux density limit at an angular distance $\theta$ and, similarly, the UU for the un-lensed case.

The results for two representative cases are compared with the measured cross-correlation signal in Fig. \ref{fig:strlen}. If we adopt, as the foreground typical lens an halo mass of $log_{10}(M/M_\odot)=13.06$ (the $M_{min}^{lens}$, determined from the default foreground sample), and we do not introduce any constraint on the gravitational lensing effect, we obtain the cyan dot-dot-dashed line. It is reassuring that, for a relatively low mass deflector, the simulation signal is very similar to the one expected from the 1-halo term under the weak lensing approach for the same typical $M_{min}$ deflector mass. Even in this case, there is a non-negligible strong lensing effect that increases slightly the correlation strength below $\sim 30"$. On the other hand, when we consider a much massive deflector, $log_{10}(M/M_\odot)=13.9$, we have a correlation signal similar to the observed one. Notice that if we don't limit the amplification (i.e. consider only the strong lensing effect or $\mu>1.7$ for this particular mass deflector), we obtain a clear overestimation  of the correlation at larger angular scales due to the weak lensing contribution of such massive halos. In this case it is clear that assuming that all $\sim10^5$ deflectors have $log_{10}(M/M_\odot)=13.9$ is not physically realistic, but it is useful to demonstrate that the enhancement of the measured signal at these small scales is easily explained by the proper modeling of the strong gravitational effect and the statistical characteristics of the deflectors.

The detailed modeling of the cross-correlation signal at these scales (mass function, mas density profiles, pointing uncertainties, etc) is beyond the scope of this paper, and will constitute the main scope of a future publication. 

\section{Conclusions}

In this work we measure and study the cross-correlation signal between a foreground sample of GAMA galaxies with spectroscopic redshifts in the range $0.2<z<0.8$, and a background sample of H-ATLAS galaxies with $z>1.2$. It constitutes a substantial improvement over the GN14 cross-correlation measurements with newer catalogues, more surveyed area and a sample selection that improves the statistics and, as a consequence, leads to smaller uncertainties (with $S/N\gtrsim 5$ below 10 arcmin and reaching $S/N\sim 20$ below 30 arcsec).

Thanks to the background sample source number count steepness, $\beta\sim3$, we are able to measure the same signal by splitting the default foreground sample in three different redshift bins ($0.2<z<0.3$, $0.3<z<0.5$ and $0.5<z<0.8$ ). With the addition of another redshift bin at lower redshift ($0.1<z<0.2$), we are able to perform a tomographic analysis of the magnification bias. We achieve measurements in all the redshift bins with $S/N\gtrsim 3$ below 10 arcmin and reaching $S/N\sim 15$ below 30 arcsec.

In the case of the background sources we apply a traditional halo model to their auto-correlation function finding the typical HOD masses values in agreement with previous studies: a minimum halo mass to host a central galaxy, $M_{min}\sim 10^{12.26} M_\odot$, and a pivot halo mass to have at least one sub-halo satellite, $M_1\sim 10^{12.84} M_\odot$. 

The halo modeling of the measured cross-correlation signals allow us to study also the typical mass of the foreground deflectors involved and, thanks to the tomographic analysis, to determine it variation with redshift.  As summarized in Table \ref{tab:bfv}, the best fit values determined from our main foreground sample are (mean and 68\% confidence intervals): $log_{10}(M_{min}^{lens}/M_\odot)=13.06^{+0.05}_{-0.06}$, $log_{10}(M_1^{lens}/M_\odot)=14.57^{+0.22}_{-0.16}$ and $\alpha_{sat}^{lens}=2.92^{+1.12}_{-0.78}$. However, the $\alpha_{sat}^{lens}$ parameter is almost unconstrained and it derived value depends directly on the imposed flat priors.
These results confirm the GN14 conclusions that the signal is mainly produced by very massive foreground galaxies or even galaxy groups/clusters that are signposted by those. The tomographic analysis shows that, while $M_1^{lens}$ and $\alpha_{sat}^{lens}$ are almost redshift independent, there is a clear evolution of an increasing $M_{min}^{lens}$ values with redshift. This evolution is mainly reflecting the possibility to increase the deflector mass (i.e. to reduce the number of potential deflectors) thanks to the higher weak gravitational lensing probability at higher foreground redshifts.

Finally, the halo modeling was also useful to point out the typical angular scale where the strong lensing can not be considered negligible any more at $\sim 30"$ --- an starting assumption in the theoretical model. This interpretation is supported by the results of basic but effective simulations.

This paper constitutes the first of a series of works that will investigate in detail the different aspect of the measured cross-correlation function in order to study the diverse kind of astrophysical and cosmological constraints that can be extracted from it.

\acknowledgments

We thank the anonymous referee for the useful comments that helped in improving the original paper. We also thank M.J. Micha{\l}owski and A. Amvrosiadis for useful and insightful comments.

The authors acknowledge financial support from the I+D 2015 project AYA2015-65887-P (MINECO/FEDER). J.G.N also acknowledges financial support from the Spanish MINECO for a ``Ramon y Cajal'' fellowship (RYC-2013-13256). AL acknowledges partial support by PRIN MIUR 2015 `Cosmology
and Fundamental Physics: illuminating the Dark Universe with Euclid', PRIN INAF 2014 `Probing the AGN/galaxy co-evolution through ultra-deep and ultra-high-resolution radio surveys', and the RADIOFOREGROUNDS grant (COMPET-05-2015, agreement number 687312) of the European Union Horizon 2020 research and innovation programme. LD, RJI and SJM acknowledge support from the ERC Advanced Grant COSMICISM, and LD and SJM also acknowledge support from the ERC Consolidator Grant, CosmicDust. MN  received funding from the European Union's Horizon 2020 research and innovation programme under the Marie Sk{\l}odowska-Curie grant agreement No 707601. GDZ acknowledges support from ASI/INAF agreement n.~2014-024-R.1 for the {\it Planck} LFI Activity of Phase E2 and from the ASI/Physics Department of the university of Roma--Tor Vergata for Study activities of the italian cosmology community n. 2016-24-H.0.

The Herschel-ATLAS is a project with Herschel, which is an ESA space observatory with science instruments provided by European-led Principal Investigator consortia and with important participation from NASA. The H-ATLAS website is http://www.h-atlas.org/
GAMA is a joint European-Australasian project based around a spectroscopic campaign using the Anglo- Australian Telescope. The GAMA input catalogue is based on data taken from the Sloan Digital Sky Survey and the UKIRT Infrared Deep Sky Survey. Complementary imaging of the GAMA regions is being obtained by a number of independent survey programs including GALEX MIS, VST KIDS, VISTA VIKING, WISE, Herschel-ATLAS, GMRT and ASKAP providing UV to radio coverage. GAMA is funded by the STFC (UK), the ARC (Australia), the AAO, and the participating institutions. The GAMA website is: http://www.gama-survey.org/

The simulations were carried out at the `Centro Interuniversitario del Nord-Est per il Calcolo Elettronico' (CINECA, Bologna), with CPU time assigned under the project Sis16\_COSMOGAL, ref. tts\#345223.

This research has made use of \texttt{TopCat} \cite{topcat}, the Ned Wright's Cosmology Calculator \cite{nedcc} and the python packages \texttt{PyMC}, \texttt{corner.py} \cite{corner} and \texttt{Astropy}, a community-developed core Python package for Astronomy \cite{astropy}.


\bibliography{./xcorr_astro}{}

\providecommand{\href}[2]{#2}\begingroup\raggedright\begin{thebibliography}{10}

\bibitem{Sch92}
P.~{Schneider}, J.~{Ehlers} and E.~E. {Falco}, \emph{{Gravitational Lenses}}.
\newblock 1992,
  \href{http://dx.doi.org/10.1007/978-3-662-03758-4}{10.1007/978-3-662-03758-4}.

\bibitem{Bar93}
M.~{Bartelmann} and P.~{Schneider}, \emph{{Large-scale correlations between
  QSOs and galaxies - an effect caused by gravitational lensing?}},
  {\emph{\aap} {\bf 268} (Feb., 1993) 1--13}.

\bibitem{Moe98}
R.~{Moessner} and B.~{Jain}, \emph{{Angular cross-correlation of galaxies - A
  probe of gravitational lensing by large-scale structure}},
  \href{http://dx.doi.org/10.1046/j.1365-8711.1998.01378.x}{\emph{\mnras} {\bf
  294} (Feb., 1998) L18--L24},
  [\href{https://arxiv.org/abs/arXiv:astro-ph/9709159}{{\tt
  arXiv:astro-ph/9709159}}].

\bibitem{Scr05}
R.~{Scranton}, B.~{M{\'e}nard}, G.~T. {Richards}, R.~C. {Nichol}, A.~D.
  {Myers}, B.~{Jain} et~al., \emph{{Detection of Cosmic Magnification with the
  Sloan Digital Sky Survey}},
  \href{http://dx.doi.org/10.1086/431358}{\emph{\apj} {\bf 633} (Nov., 2005)
  589--602}, [\href{https://arxiv.org/abs/arXiv:astro-ph/0504510}{{\tt
  arXiv:astro-ph/0504510}}].

\bibitem{Men10}
B.~{M{\'e}nard}, R.~{Scranton}, M.~{Fukugita} and G.~{Richards},
  \emph{{Measuring the galaxy-mass and galaxy-dust correlations through
  magnification and reddening}},
  \href{http://dx.doi.org/10.1111/j.1365-2966.2010.16486.x}{\emph{\mnras} {\bf
  405} (June, 2010) 1025--1039}, [\href{https://arxiv.org/abs/0902.4240}{{\tt
  0902.4240}}].

\bibitem{Hil13}
H.~{Hildebrandt}, L.~{van Waerbeke}, D.~{Scott}, M.~{B{\'e}thermin}, J.~{Bock},
  D.~{Clements} et~al., \emph{{Inferring the mass of submillimetre galaxies by
  exploiting their gravitational magnification of background galaxies}},
  \href{http://dx.doi.org/10.1093/mnras/sts585}{\emph{\mnras} {\bf 429} (Mar.,
  2013) 3230--3237}, [\href{https://arxiv.org/abs/1212.2650}{{\tt 1212.2650}}].

\bibitem{Bar01}
M.~{Bartelmann} and P.~{Schneider}, \emph{{Weak gravitational lensing}},
  \href{http://dx.doi.org/10.1016/S0370-1573(00)00082-X}{\emph{Phys. Rep.} {\bf
  340} (Jan., 2001) 291--472},
  [\href{https://arxiv.org/abs/arXiv:astro-ph/9912508}{{\tt
  arXiv:astro-ph/9912508}}].

\bibitem{Alm05}
O.~{Almaini}, J.~S. {Dunlop}, C.~J. {Willott}, D.~M. {Alexander}, F.~E. {Bauer}
  and C.~T. {Liu}, \emph{{Correlations between bright submillimetre sources and
  low-redshift galaxies}},
  \href{http://dx.doi.org/10.1111/j.1365-2966.2005.08790.x}{\emph{\mnras} {\bf
  358} (Apr., 2005) 875--882},
  [\href{https://arxiv.org/abs/astro-ph/0501169}{{\tt astro-ph/0501169}}].

\bibitem{Bla06}
C.~{Blake}, A.~{Pope}, D.~{Scott} and B.~{Mobasher}, \emph{{On the
  cross-correlation of sub-mm sources and optically selected galaxies}},
  \href{http://dx.doi.org/10.1111/j.1365-2966.2006.10158.x}{\emph{\mnras} {\bf
  368} (May, 2006) 732--740},
  [\href{https://arxiv.org/abs/astro-ph/0602428}{{\tt astro-ph/0602428}}].

\bibitem{Pil10}
G.~L. {Pilbratt}, J.~R. {Riedinger}, T.~{Passvogel}, G.~{Crone}, D.~{Doyle},
  U.~{Gageur} et~al., \emph{{Herschel Space Observatory. An ESA facility for
  far-infrared and submillimetre astronomy}},
  \href{http://dx.doi.org/10.1051/0004-6361/201014759}{\emph{\aap} {\bf 518}
  (July, 2010) L1}, [\href{https://arxiv.org/abs/1005.5331}{{\tt 1005.5331}}].

\bibitem{Eal10}
S.~{Eales}, L.~{Dunne}, D.~{Clements}, A.~{Cooray}, G.~{de Zotti}, S.~{Dye}
  et~al., \emph{{The Herschel ATLAS}},
  \href{http://dx.doi.org/10.1086/653086}{\emph{\pasp} {\bf 122} (May, 2010)
  499--515}, [\href{https://arxiv.org/abs/0910.4279}{{\tt 0910.4279}}].

\bibitem{Neg10}
M.~{Negrello}, R.~{Hopwood}, G.~{De Zotti}, A.~{Cooray}, A.~{Verma}, J.~{Bock}
  et~al., \emph{{The Detection of a Population of Submillimeter-Bright,
  Strongly Lensed Galaxies}},
  \href{http://dx.doi.org/10.1126/science.1193420}{\emph{Science} {\bf 330}
  (Nov., 2010) 800}, [\href{https://arxiv.org/abs/1011.1255}{{\tt 1011.1255}}].

\bibitem{GN12}
J.~{Gonz{\'a}lez-Nuevo}, A.~{Lapi}, S.~{Fleuren}, S.~{Bressan}, L.~{Danese},
  G.~{De Zotti} et~al., \emph{{Herschel-ATLAS: Toward a Sample of \~{}1000
  Strongly Lensed Galaxies}},
  \href{http://dx.doi.org/10.1088/0004-637X/749/1/65}{\emph{\apj} {\bf 749}
  (Apr., 2012) 65}, [\href{https://arxiv.org/abs/1202.0402}{{\tt 1202.0402}}].

\bibitem{Neg14}
M.~{Negrello}, R.~{Hopwood}, S.~{Dye}, E.~d. {Cunha}, S.~{Serjeant}, J.~{Fritz}
  et~al., \emph{{Herschel *-ATLAS: deep HST/WFC3 imaging of strongly lensed
  submillimetre galaxies}},
  \href{http://dx.doi.org/10.1093/mnras/stu413}{\emph{\mnras} {\bf 440} (May,
  2014) 1999--2012}, [\href{https://arxiv.org/abs/1311.5898}{{\tt 1311.5898}}].

\bibitem{Neg17}
M.~{Negrello}, S.~{Amber}, A.~{Amvrosiadis}, Z.-Y. {Cai}, A.~{Lapi},
  J.~{Gonzalez-Nuevo} et~al., \emph{{The Herschel-ATLAS: a sample of 500
  {$\mu$}m-selected lensed galaxies over 600 deg$^{2}$}},
  \href{http://dx.doi.org/10.1093/mnras/stw2911}{\emph{\mnras} {\bf 465} (Mar.,
  2017) 3558--3580}, [\href{https://arxiv.org/abs/1611.03922}{{\tt
  1611.03922}}].

\bibitem{Fu12}
H.~{Fu}, E.~{Jullo}, A.~{Cooray}, R.~S. {Bussmann}, R.~J. {Ivison},
  I.~{P{\'e}rez-Fournon} et~al., \emph{{A Comprehensive View of a Strongly
  Lensed Planck-Associated Submillimeter Galaxy}},
  \href{http://dx.doi.org/10.1088/0004-637X/753/2/134}{\emph{\apj} {\bf 753}
  (July, 2012) 134}, [\href{https://arxiv.org/abs/1202.1829}{{\tt 1202.1829}}].

\bibitem{Bus12}
R.~S. {Bussmann}, M.~A. {Gurwell}, H.~{Fu}, D.~J.~B. {Smith}, S.~{Dye},
  R.~{Auld} et~al., \emph{{A Detailed Gravitational Lens Model Based on
  Submillimeter Array and Keck Adaptive Optics Imaging of a Herschel-ATLAS
  Submillimeter Galaxy at z = 4.243}},
  \href{http://dx.doi.org/10.1088/0004-637X/756/2/134}{\emph{\apj} {\bf 756}
  (Sept., 2012) 134}, [\href{https://arxiv.org/abs/1207.2724}{{\tt
  1207.2724}}].

\bibitem{Bus13}
R.~S. {Bussmann}, I.~{P{\'e}rez-Fournon}, S.~{Amber}, J.~{Calanog}, M.~A.
  {Gurwell}, H.~{Dannerbauer} et~al., \emph{{Gravitational Lens Models Based on
  Submillimeter Array Imaging of Herschel-selected Strongly Lensed
  Sub-millimeter Galaxies at z>1.5}},
  \href{http://dx.doi.org/10.1088/0004-637X/779/1/25}{\emph{\apj} {\bf 779}
  (Dec., 2013) 25}, [\href{https://arxiv.org/abs/1309.0836}{{\tt 1309.0836}}].

\bibitem{War13}
J.~L. {Wardlow}, A.~{Cooray}, F.~{De Bernardis}, A.~{Amblard}, V.~{Arumugam},
  H.~{Aussel} et~al., \emph{{HerMES: Candidate Gravitationally Lensed Galaxies
  and Lensing Statistics at Submillimeter Wavelengths}},
  \href{http://dx.doi.org/10.1088/0004-637X/762/1/59}{\emph{\apj} {\bf 762}
  (Jan., 2013) 59}, [\href{https://arxiv.org/abs/1205.3778}{{\tt 1205.3778}}].

\bibitem{Cal14}
J.~A. {Calanog}, H.~{Fu}, A.~{Cooray}, J.~{Wardlow}, B.~{Ma}, S.~{Amber}
  et~al., \emph{{Lens Models of Herschel-selected Galaxies from High-resolution
  Near-IR Observations}},
  \href{http://dx.doi.org/10.1088/0004-637X/797/2/138}{\emph{\apj} {\bf 797}
  (Dec., 2014) 138}, [\href{https://arxiv.org/abs/1406.1487}{{\tt 1406.1487}}].

\bibitem{Nay16}
H.~{Nayyeri}, M.~{Keele}, A.~{Cooray}, D.~A. {Riechers}, R.~J. {Ivison}, A.~I.
  {Harris} et~al., \emph{{Candidate Gravitationally Lensed Dusty Star-forming
  Galaxies in the Herschel Wide Area Surveys}},
  \href{http://dx.doi.org/10.3847/0004-637X/823/1/17}{\emph{\apj} {\bf 823}
  (May, 2016) 17}, [\href{https://arxiv.org/abs/1601.03401}{{\tt 1601.03401}}].

\bibitem{Vie13}
J.~D. {Vieira}, D.~P. {Marrone}, S.~C. {Chapman}, C.~{De Breuck}, Y.~D.
  {Hezaveh}, A.~{Wei{$\beta$}} et~al., \emph{{Dusty starburst galaxies in the
  early Universe as revealed by gravitational lensing}},
  \href{http://dx.doi.org/10.1038/nature12001}{\emph{Nature} {\bf 495} (Mar.,
  2013) 344--347}, [\href{https://arxiv.org/abs/1303.2723}{{\tt 1303.2723}}].

\bibitem{Spi16}
J.~S. {Spilker}, D.~P. {Marrone}, M.~{Aravena}, M.~{B{\'e}thermin}, M.~S.
  {Bothwell}, J.~E. {Carlstrom} et~al., \emph{{ALMA Imaging and Gravitational
  Lens Models of South Pole Telescope:Selected Dusty, Star-Forming Galaxies at
  High Redshifts}},
  \href{http://dx.doi.org/10.3847/0004-637X/826/2/112}{\emph{\apj} {\bf 826}
  (Aug., 2016) 112}, [\href{https://arxiv.org/abs/1604.05723}{{\tt
  1604.05723}}].

\bibitem{PIPXXVII}
{Planck Collaboration}, N.~{Aghanim}, B.~{Altieri}, M.~{Arnaud}, M.~{Ashdown},
  J.~{Aumont} et~al., \emph{{Planck intermediate results. XXVII. High-redshift
  infrared galaxy overdensity candidates and lensed sources discovered by
  Planck and confirmed by Herschel-SPIRE}},
  \href{http://dx.doi.org/10.1051/0004-6361/201424790}{\emph{\aap} {\bf 582}
  (Oct., 2015) A30}, [\href{https://arxiv.org/abs/1503.08773}{{\tt
  1503.08773}}].

\bibitem{Can15}
R.~{Ca{\~n}ameras}, N.~P.~H. {Nesvadba}, D.~{Guery}, T.~{McKenzie},
  S.~{K{\"o}nig}, G.~{Petitpas} et~al., \emph{{Planck's dusty GEMS: The
  brightest gravitationally lensed galaxies discovered with the Planck all-sky
  survey}}, \href{http://dx.doi.org/10.1051/0004-6361/201425128}{\emph{\aap}
  {\bf 581} (Sept., 2015) A105}, [\href{https://arxiv.org/abs/1506.01962}{{\tt
  1506.01962}}].

\bibitem{Har16}
K.~C. {Harrington}, M.~S. {Yun}, R.~{Cybulski}, G.~W. {Wilson}, I.~{Aretxaga},
  M.~{Chavez} et~al., \emph{{Early science with the Large Millimeter Telescope:
  observations of extremely luminous high-z sources identified by Planck}},
  \href{http://dx.doi.org/10.1093/mnras/stw614}{\emph{\mnras} {\bf 458} (June,
  2016) 4383--4399}, [\href{https://arxiv.org/abs/1603.05622}{{\tt
  1603.05622}}].

\bibitem{Nes16}
N.~{Nesvadba}, R.~{Kneissl}, R.~{Ca{\~n}ameras}, F.~{Boone}, E.~{Falgarone},
  B.~{Frye} et~al., \emph{{Planck's Dusty GEMS. II. Extended [CII] emission and
  absorption in the Garnet at z = 3.4 seen with ALMA}},
  \href{http://dx.doi.org/10.1051/0004-6361/201629037}{\emph{\aap} {\bf 593}
  (Aug., 2016) L2}, [\href{https://arxiv.org/abs/1610.01169}{{\tt
  1610.01169}}].

\bibitem{Can17}
R.~{Ca{\~n}ameras}, N.~P.~H. {Nesvadba}, R.~{Kneissl}, M.~{Limousin},
  R.~{Gavazzi}, D.~{Scott} et~al., \emph{{Planck's dusty GEMS. III. A massive
  lensing galaxy with a bottom-heavy stellar initial mass function at z =
  1.5}}, \href{http://dx.doi.org/10.1051/0004-6361/201630359}{\emph{\aap} {\bf
  600} (Apr., 2017) L3}, [\href{https://arxiv.org/abs/1703.02984}{{\tt
  1703.02984}}].

\bibitem{Bou14}
N.~{Bourne}, S.~J. {Maddox}, L.~{Dunne}, S.~{Dye}, S.~{Eales}, C.~{Hoyos}
  et~al., \emph{{Colour matters: the effects of lensing on the positional
  offsets between optical and submillimetre galaxies in Herschel-ATLAS}},
  \href{http://dx.doi.org/10.1093/mnras/stu1582}{\emph{\mnras} {\bf 444} (Oct.,
  2014) 1884--1892}, [\href{https://arxiv.org/abs/1407.5994}{{\tt 1407.5994}}].

\bibitem{Wan11}
L.~{Wang}, A.~{Cooray}, D.~{Farrah}, A.~{Amblard}, R.~{Auld}, J.~{Bock} et~al.,
  \emph{{HerMES: detection of cosmic magnification of submillimetre galaxies
  using angular cross-correlation}},
  \href{http://dx.doi.org/10.1111/j.1365-2966.2011.18417.x}{\emph{\mnras} {\bf
  414} (June, 2011) 596--601}, [\href{https://arxiv.org/abs/1101.4796}{{\tt
  1101.4796}}].

\bibitem{GN14}
J.~{Gonz{\'a}lez-Nuevo}, A.~{Lapi}, M.~{Negrello}, L.~{Danese}, G.~{De Zotti},
  S.~{Amber} et~al., \emph{{Herschel-ATLAS/GAMA: SDSS cross-correlation induced
  by weak lensing}},
  \href{http://dx.doi.org/10.1093/mnras/stu1041}{\emph{\mnras} {\bf 442} (Aug.,
  2014) 2680--2690}, [\href{https://arxiv.org/abs/1401.4094}{{\tt 1401.4094}}].

\bibitem{Ahn12}
C.~P. {Ahn}, R.~{Alexandroff}, C.~{Allende Prieto}, S.~F. {Anderson},
  T.~{Anderton}, B.~H. {Andrews} et~al., \emph{{The Ninth Data Release of the
  Sloan Digital Sky Survey: First Spectroscopic Data from the SDSS-III Baryon
  Oscillation Spectroscopic Survey}},
  \href{http://dx.doi.org/10.1088/0067-0049/203/2/21}{\emph{\apjs} {\bf 203}
  (Dec., 2012) 21}, [\href{https://arxiv.org/abs/1207.7137}{{\tt 1207.7137}}].

\bibitem{Dri11}
S.~P. {Driver}, D.~T. {Hill}, L.~S. {Kelvin}, A.~S.~G. {Robotham}, J.~{Liske},
  P.~{Norberg} et~al., \emph{{Galaxy and Mass Assembly (GAMA): survey
  diagnostics and core data release}},
  \href{http://dx.doi.org/10.1111/j.1365-2966.2010.18188.x}{\emph{\mnras} {\bf
  413} (May, 2011) 971--995}, [\href{https://arxiv.org/abs/1009.0614}{{\tt
  1009.0614}}].

\bibitem{Coo02}
A.~{Cooray} and R.~{Sheth}, \emph{{Halo models of large scale structure}},
  \href{http://dx.doi.org/10.1016/S0370-1573(02)00276-4}{\emph{Phys. Rep.} {\bf
  372} (Dec., 2002) 1--129},
  [\href{https://arxiv.org/abs/astro-ph/0206508}{{\tt astro-ph/0206508}}].

\bibitem{Sel00}
U.~{Seljak}, \emph{{Analytic model for galaxy and dark matter clustering}},
  \href{http://dx.doi.org/10.1046/j.1365-8711.2000.03715.x}{\emph{\mnras} {\bf
  318} (Oct., 2000) 203--213},
  [\href{https://arxiv.org/abs/astro-ph/0001493}{{\tt astro-ph/0001493}}].

\bibitem{Guz01}
J.~{Guzik} and U.~{Seljak}, \emph{{Galaxy-dark matter correlations applied to
  galaxy-galaxy lensing: predictions from the semi-analytic galaxy formation
  models}},
  \href{http://dx.doi.org/10.1046/j.1365-8711.2001.04081.x}{\emph{\mnras} {\bf
  321} (Mar., 2001) 439--449},
  [\href{https://arxiv.org/abs/astro-ph/0007067}{{\tt astro-ph/0007067}}].

\bibitem{Set99}
R.~K. {Sheth} and G.~{Tormen}, \emph{{Large-scale bias and the peak background
  split}},
  \href{http://dx.doi.org/10.1046/j.1365-8711.1999.02692.x}{\emph{\mnras} {\bf
  308} (Sept., 1999) 119--126},
  [\href{https://arxiv.org/abs/astro-ph/9901122}{{\tt astro-ph/9901122}}].

\bibitem{NFW96}
J.~F. {Navarro}, C.~S. {Frenk} and S.~D.~M. {White}, \emph{{The Structure of
  Cold Dark Matter Halos}}, \href{http://dx.doi.org/10.1086/177173}{\emph{\apj}
  {\bf 462} (May, 1996) 563},
  [\href{https://arxiv.org/abs/astro-ph/9508025}{{\tt astro-ph/9508025}}].

\bibitem{Bul01}
J.~S. {Bullock}, T.~S. {Kolatt}, Y.~{Sigad}, R.~S. {Somerville}, A.~V.
  {Kravtsov}, A.~A. {Klypin} et~al., \emph{{Profiles of dark haloes: evolution,
  scatter and environment}},
  \href{http://dx.doi.org/10.1046/j.1365-8711.2001.04068.x}{\emph{\mnras} {\bf
  321} (Mar., 2001) 559--575},
  [\href{https://arxiv.org/abs/astro-ph/9908159}{{\tt astro-ph/9908159}}].

\bibitem{Zen05}
Z.~{Zheng}, A.~A. {Berlind}, D.~H. {Weinberg}, A.~J. {Benson}, C.~M. {Baugh},
  S.~{Cole} et~al., \emph{{Theoretical Models of the Halo Occupation
  Distribution: Separating Central and Satellite Galaxies}},
  \href{http://dx.doi.org/10.1086/466510}{\emph{\apj} {\bf 633} (Nov., 2005)
  791--809}, [\href{https://arxiv.org/abs/astro-ph/0408564}{{\tt
  astro-ph/0408564}}].

\bibitem{Lim53}
D.~N. {Limber}, \emph{{The Analysis of Counts of the Extragalactic Nebulae in
  Terms of a Fluctuating Density Field.}},
  \href{http://dx.doi.org/10.1086/145672}{\emph{\apj} {\bf 117} (Jan., 1953)
  134}.

\bibitem{Kil17}
M.~{Kilbinger}, C.~{Heymans}, M.~{Asgari}, S.~{Joudaki}, P.~{Schneider},
  P.~{Simon} et~al., \emph{{Precision calculations of the cosmic shear power
  spectrum projection}}, {\emph{ArXiv e-prints} (Feb., 2017) },
  [\href{https://arxiv.org/abs/1702.05301}{{\tt 1702.05301}}].

\bibitem{Pog10}
A.~{Poglitsch}, C.~{Waelkens}, N.~{Geis}, H.~{Feuchtgruber},
  B.~{Vandenbussche}, L.~{Rodriguez} et~al., \emph{{The Photodetector Array
  Camera and Spectrometer (PACS) on the Herschel Space Observatory}},
  \href{http://dx.doi.org/10.1051/0004-6361/201014535}{\emph{\aap} {\bf 518}
  (July, 2010) L2}, [\href{https://arxiv.org/abs/1005.1487}{{\tt 1005.1487}}].

\bibitem{Gri10}
M.~J. {Griffin}, A.~{Abergel}, A.~{Abreu}, P.~A.~R. {Ade}, P.~{Andr{\'e}},
  J.-L. {Augueres} et~al., \emph{{The Herschel-SPIRE instrument and its
  in-flight performance}},
  \href{http://dx.doi.org/10.1051/0004-6361/201014519}{\emph{\aap} {\bf 518}
  (July, 2010) L3}, [\href{https://arxiv.org/abs/1005.5123}{{\tt 1005.5123}}].

\bibitem{Iba10}
E.~{Ibar}, R.~J. {Ivison}, A.~{Cava}, G.~{Rodighiero}, S.~{Buttiglione},
  P.~{Temi} et~al., \emph{{H-ATLAS: PACS imaging for the Science Demonstration
  Phase}},
  \href{http://dx.doi.org/10.1111/j.1365-2966.2010.17620.x}{\emph{\mnras} {\bf
  409} (Nov., 2010) 38--47}, [\href{https://arxiv.org/abs/1009.0262}{{\tt
  1009.0262}}].

\bibitem{Pas11}
E.~{Pascale}, R.~{Auld}, A.~{Dariush}, L.~{Dunne}, S.~{Eales}, S.~{Maddox}
  et~al., \emph{{The first release of data from the Herschel ATLAS: the SPIRE
  images}},
  \href{http://dx.doi.org/10.1111/j.1365-2966.2011.18756.x}{\emph{\mnras} {\bf
  415} (July, 2011) 911--917}, [\href{https://arxiv.org/abs/1010.5782}{{\tt
  1010.5782}}].

\bibitem{Rig11}
E.~E. {Rigby}, S.~J. {Maddox}, L.~{Dunne}, M.~{Negrello}, D.~J.~B. {Smith},
  J.~{Gonz{\'a}lez-Nuevo} et~al., \emph{{Herschel-ATLAS: first data release of
  the Science Demonstration Phase source catalogues}},
  \href{http://dx.doi.org/10.1111/j.1365-2966.2011.18864.x}{\emph{\mnras} {\bf
  415} (Aug., 2011) 2336--2348}, [\href{https://arxiv.org/abs/1010.5787}{{\tt
  1010.5787}}].

\bibitem{Val16}
E.~{Valiante}, M.~W.~L. {Smith}, S.~{Eales}, S.~J. {Maddox}, E.~{Ibar},
  R.~{Hopwood} et~al., \emph{{The Herschel-ATLAS data release 1 - I. Maps,
  catalogues and number counts}},
  \href{http://dx.doi.org/10.1093/mnras/stw1806}{\emph{\mnras} {\bf 462} (Nov.,
  2016) 3146--3179}, [\href{https://arxiv.org/abs/1606.09615}{{\tt
  1606.09615}}].

\bibitem{Bou16}
N.~{Bourne}, L.~{Dunne}, S.~J. {Maddox}, S.~{Dye}, C.~{Furlanetto}, C.~{Hoyos}
  et~al., \emph{{The Herschel-ATLAS Data Release 1 - II. Multi-wavelength
  counterparts to submillimetre sources}},
  \href{http://dx.doi.org/10.1093/mnras/stw1654}{\emph{\mnras} {\bf 462} (Oct.,
  2016) 1714--1734}, [\href{https://arxiv.org/abs/1606.09254}{{\tt
  1606.09254}}].

\bibitem{Lap11}
A.~{Lapi}, J.~{Gonz{\'a}lez-Nuevo}, L.~{Fan}, A.~{Bressan}, G.~{De Zotti},
  L.~{Danese} et~al., \emph{{Herschel-ATLAS Galaxy Counts and High-redshift
  Luminosity Functions: The Formation of Massive Early-type Galaxies}},
  \href{http://dx.doi.org/10.1088/0004-637X/742/1/24}{\emph{\apj} {\bf 742}
  (Nov., 2011) 24}, [\href{https://arxiv.org/abs/1108.3911}{{\tt 1108.3911}}].

\bibitem{Ivi10}
R.~J. {Ivison}, A.~M. {Swinbank}, B.~{Swinyard}, I.~{Smail}, C.~P. {Pearson},
  D.~{Rigopoulou} et~al., \emph{{Herschel and SCUBA-2 imaging and spectroscopy
  of a bright, lensed submillimetre galaxy at z = 2.3}},
  \href{http://dx.doi.org/10.1051/0004-6361/201014548}{\emph{\aap} {\bf 518}
  (July, 2010) L35}, [\href{https://arxiv.org/abs/1005.1071}{{\tt 1005.1071}}].

\bibitem{Swi10}
A.~M. {Swinbank}, I.~{Smail}, S.~{Longmore}, A.~I. {Harris}, A.~J. {Baker},
  C.~{De Breuck} et~al., \emph{{Intense star formation within resolved compact
  regions in a galaxy at z = 2.3}},
  \href{http://dx.doi.org/10.1038/nature08880}{\emph{Nature} {\bf 464} (Apr.,
  2010) 733--736}, [\href{https://arxiv.org/abs/1003.3674}{{\tt 1003.3674}}].

\bibitem{Pea13}
E.~A. {Pearson}, S.~{Eales}, L.~{Dunne}, J.~{Gonzalez-Nuevo}, S.~{Maddox},
  J.~E. {Aguirre} et~al., \emph{{H-ATLAS: estimating redshifts of Herschel
  sources from sub-mm fluxes}},
  \href{http://dx.doi.org/10.1093/mnras/stt1369}{\emph{\mnras} (Sept., 2013) },
  [\href{https://arxiv.org/abs/1308.5681}{{\tt 1308.5681}}].

\bibitem{Ivi16}
R.~J. {Ivison}, A.~J.~R. {Lewis}, A.~{Weiss}, V.~{Arumugam}, J.~M. {Simpson},
  W.~S. {Holland} et~al., \emph{{The Space Density of Luminous Dusty
  Star-forming Galaxies at z>4: SCUBA-2 and LABOCA Imaging of Ultrared Galaxies
  from Herschel-ATLAS}},
  \href{http://dx.doi.org/10.3847/0004-637X/832/1/78}{\emph{\apj} {\bf 832}
  (Nov., 2016) 78}, [\href{https://arxiv.org/abs/1611.00762}{{\tt
  1611.00762}}].

\bibitem{Rie13}
D.~A. {Riechers}, C.~M. {Bradford}, D.~L. {Clements}, C.~D. {Dowell},
  I.~{P{\'e}rez-Fournon}, R.~J. {Ivison} et~al., \emph{{A dust-obscured massive
  maximum-starburst galaxy at a redshift of 6.34}},
  \href{http://dx.doi.org/10.1038/nature12050}{\emph{Nature} {\bf 496} (Apr.,
  2013) 329--333}, [\href{https://arxiv.org/abs/1304.4256}{{\tt 1304.4256}}].

\bibitem{Wei13}
A.~{Wei{\ss}}, C.~{De Breuck}, D.~P. {Marrone}, J.~D. {Vieira}, J.~E.
  {Aguirre}, K.~A. {Aird} et~al., \emph{{ALMA Redshifts of Millimeter-selected
  Galaxies from the SPT Survey: The Redshift Distribution of Dusty Star-forming
  Galaxies}}, \href{http://dx.doi.org/10.1088/0004-637X/767/1/88}{\emph{\apj}
  {\bf 767} (Apr., 2013) 88}, [\href{https://arxiv.org/abs/1303.2726}{{\tt
  1303.2726}}].

\bibitem{Asb16}
V.~{Asboth}, A.~{Conley}, J.~{Sayers}, M.~{B{\'e}thermin}, S.~C. {Chapman},
  D.~L. {Clements} et~al., \emph{{HerMES: a search for high-redshift dusty
  galaxies in the HerMES Large Mode Survey - catalogue, number counts and early
  results}}, \href{http://dx.doi.org/10.1093/mnras/stw1769}{\emph{\mnras} {\bf
  462} (Oct., 2016) 1989--2000}, [\href{https://arxiv.org/abs/1601.02665}{{\tt
  1601.02665}}].

\bibitem{Str16}
M.~L. {Strandet}, A.~{Weiss}, J.~D. {Vieira}, C.~{de Breuck}, J.~E. {Aguirre},
  M.~{Aravena} et~al., \emph{{The Redshift Distribution of Dusty Star-forming
  Galaxies from the SPT Survey}},
  \href{http://dx.doi.org/10.3847/0004-637X/822/2/80}{\emph{\apj} {\bf 822}
  (May, 2016) 80}, [\href{https://arxiv.org/abs/1603.05094}{{\tt 1603.05094}}].

\bibitem{Bud03}
T.~{Budav{\'a}ri}, A.~J. {Connolly}, A.~S. {Szalay}, I.~{Szapudi}, I.~{Csabai},
  R.~{Scranton} et~al., \emph{{Angular Clustering with Photometric Redshifts in
  the Sloan Digital Sky Survey: Bimodality in the Clustering Properties of
  Galaxies}}, \href{http://dx.doi.org/10.1086/377168}{\emph{\apj} {\bf 595}
  (Sept., 2003) 59--70}, [\href{https://arxiv.org/abs/astro-ph/0305603}{{\tt
  astro-ph/0305603}}].

\bibitem{Bia16}
F.~{Bianchini}, A.~{Lapi}, M.~{Calabrese}, P.~{Bielewicz}, J.~{Gonzalez-Nuevo},
  C.~{Baccigalupi} et~al., \emph{{Toward a Tomographic Analysis of the
  Cross-Correlation between Planck CMB Lensing and H-ATLAS Galaxies}},
  \href{http://dx.doi.org/10.3847/0004-637X/825/1/24}{\emph{\apj} {\bf 825}
  (July, 2016) 24}, [\href{https://arxiv.org/abs/1511.05116}{{\tt
  1511.05116}}].

\bibitem{Bal10}
I.~K. {Baldry}, A.~S.~G. {Robotham}, D.~T. {Hill}, S.~P. {Driver}, J.~{Liske},
  P.~{Norberg} et~al., \emph{{Galaxy And Mass Assembly (GAMA): the input
  catalogue and star-galaxy separation}},
  \href{http://dx.doi.org/10.1111/j.1365-2966.2010.16282.x}{\emph{\mnras} {\bf
  404} (May, 2010) 86--100}, [\href{https://arxiv.org/abs/0910.5120}{{\tt
  0910.5120}}].

\bibitem{Bal14}
I.~K. {Baldry}, M.~{Alpaslan}, A.~E. {Bauer}, J.~{Bland-Hawthorn}, S.~{Brough},
  M.~E. {Cluver} et~al., \emph{{Galaxy And Mass Assembly (GAMA): AUTOZ spectral
  redshift measurements, confidence and errors}},
  \href{http://dx.doi.org/10.1093/mnras/stu727}{\emph{\mnras} {\bf 441} (July,
  2014) 2440--2451}, [\href{https://arxiv.org/abs/1404.2626}{{\tt 1404.2626}}].

\bibitem{Lis15}
J.~{Liske}, I.~K. {Baldry}, S.~P. {Driver}, R.~J. {Tuffs}, M.~{Alpaslan},
  E.~{Andrae} et~al., \emph{{Galaxy And Mass Assembly (GAMA): end of survey
  report and data release 2}},
  \href{http://dx.doi.org/10.1093/mnras/stv1436}{\emph{\mnras} {\bf 452}
  (Sept., 2015) 2087--2126}, [\href{https://arxiv.org/abs/1506.08222}{{\tt
  1506.08222}}].

\bibitem{Con02}
A.~J. {Connolly}, R.~{Scranton}, D.~{Johnston}, S.~{Dodelson}, D.~J.
  {Eisenstein}, J.~A. {Frieman} et~al., \emph{{The Angular Correlation Function
  of Galaxies from Early Sloan Digital Sky Survey Data}},
  \href{http://dx.doi.org/10.1086/342787}{\emph{\apj} {\bf 579} (Nov., 2002)
  42--47}, [\href{https://arxiv.org/abs/arXiv:astro-ph/0107417}{{\tt
  arXiv:astro-ph/0107417}}].

\bibitem{Wan13}
Y.~{Wang}, R.~J. {Brunner} and J.~C. {Dolence}, \emph{{The SDSS galaxy angular
  two-point correlation function}},
  \href{http://dx.doi.org/10.1093/mnras/stt450}{\emph{\mnras} {\bf 432} (July,
  2013) 1961--1979}, [\href{https://arxiv.org/abs/1303.2432}{{\tt 1303.2432}}].

\bibitem{Che16b}
C.-C. {Chen}, I.~{Smail}, A.~M. {Swinbank}, J.~M. {Simpson}, O.~{Almaini},
  C.~J. {Conselice} et~al., \emph{{Faint Submillimeter Galaxies Identified
  through Their Optical/Near-infrared Colors. I. Spatial Clustering and Halo
  Masses}}, \href{http://dx.doi.org/10.3847/0004-637X/831/1/91}{\emph{\apj}
  {\bf 831} (Nov., 2016) 91}, [\href{https://arxiv.org/abs/1609.00388}{{\tt
  1609.00388}}].

\bibitem{Lan93}
S.~D. {Landy} and A.~S. {Szalay}, \emph{{Bias and variance of angular
  correlation functions}}, \href{http://dx.doi.org/10.1086/172900}{\emph{\apj}
  {\bf 412} (July, 1993) 64--71}.

\bibitem{Xia12}
J.-Q. {Xia}, M.~{Negrello}, A.~{Lapi}, G.~{De Zotti}, L.~{Danese} and
  M.~{Viel}, \emph{{Clustering of submillimetre galaxies in a self-regulated
  baryon collapse model}},
  \href{http://dx.doi.org/10.1111/j.1365-2966.2012.20705.x}{\emph{\mnras} {\bf
  422} (May, 2012) 1324--1331}, [\href{https://arxiv.org/abs/1111.4212}{{\tt
  1111.4212}}].

\bibitem{Coo10}
A.~{Cooray}, A.~{Amblard}, L.~{Wang}, V.~{Arumugam}, R.~{Auld}, H.~{Aussel}
  et~al., \emph{{HerMES: Halo occupation number and bias properties of dusty
  galaxies from angular clustering measurements}},
  \href{http://dx.doi.org/10.1051/0004-6361/201014597}{\emph{\aap} {\bf 518}
  (July, 2010) L22}, [\href{https://arxiv.org/abs/1005.3303}{{\tt 1005.3303}}].

\bibitem{Mit12}
K.~{Mitchell-Wynne}, A.~{Cooray}, Y.~{Gong}, M.~{B{\'e}thermin}, J.~{Bock},
  A.~{Franceschini} et~al., \emph{{HerMES: A Statistical Measurement of the
  Redshift Distribution of Herschel-SPIRE Sources Using the Cross-correlation
  Technique}}, \href{http://dx.doi.org/10.1088/0004-637X/753/1/23}{\emph{\apj}
  {\bf 753} (July, 2012) 23}, [\href{https://arxiv.org/abs/1203.0063}{{\tt
  1203.0063}}].

\bibitem{Wil17}
A.~{Wilkinson}, O.~{Almaini}, C.-C. {Chen}, I.~{Smail}, V.~{Arumugam},
  A.~{Blain} et~al., \emph{{The SCUBA-2 Cosmology Legacy Survey: the clustering
  of submillimetre galaxies in the UKIDSS UDS field}},
  \href{http://dx.doi.org/10.1093/mnras/stw2405}{\emph{\mnras} {\bf 464} (Jan.,
  2017) 1380--1392}, [\href{https://arxiv.org/abs/1604.00018}{{\tt
  1604.00018}}].

\bibitem{Che16a}
C.-C. {Chen}, I.~{Smail}, R.~J. {Ivison}, V.~{Arumugam}, O.~{Almaini}, C.~J.
  {Conselice} et~al., \emph{{The SCUBA-2 Cosmology Legacy Survey:
  Multiwavelength Counterparts to 10$^{3}$ Submillimeter Galaxies in the
  UKIDSS-UDS Field}},
  \href{http://dx.doi.org/10.3847/0004-637X/820/2/82}{\emph{\apj} {\bf 820}
  (Apr., 2016) 82}, [\href{https://arxiv.org/abs/1601.02630}{{\tt
  1601.02630}}].

\bibitem{Pee80}
P.~J.~E. {Peebles}, \emph{{The large-scale structure of the universe}}.
\newblock 1980.

\bibitem{Her01}
D.~{Herranz}, \emph{{Foreground-Background Galaxy Correlations in the Hubble
  Deep Fields}},  in \emph{Cosmological Physics with Gravitational Lensing}
  (J.~{Tran Thanh Van}, Y.~{Mellier} and M.~{Moniez}, eds.), p.~197, Jan.,
  2001.

\bibitem{Smi11}
D.~J.~B. {Smith}, L.~{Dunne}, S.~J. {Maddox}, S.~{Eales}, D.~G. {Bonfield},
  M.~J. {Jarvis} et~al., \emph{{Herschel-ATLAS: counterparts from the
  ultraviolet-near-infrared in the science demonstration phase catalogue}},
  \href{http://dx.doi.org/10.1111/j.1365-2966.2011.18827.x}{\emph{\mnras} {\bf
  416} (Sept., 2011) 857--872}, [\href{https://arxiv.org/abs/1007.5260}{{\tt
  1007.5260}}].

\bibitem{Bia15}
F.~{Bianchini}, P.~{Bielewicz}, A.~{Lapi}, J.~{Gonzalez-Nuevo},
  C.~{Baccigalupi}, G.~{de Zotti} et~al., \emph{{Cross-correlation between the
  CMB Lensing Potential Measured by Planck and High-z Submillimeter Galaxies
  Detected by the Herschel-Atlas Survey}},
  \href{http://dx.doi.org/10.1088/0004-637X/802/1/64}{\emph{\apj} {\bf 802}
  (Mar., 2015) 64}, [\href{https://arxiv.org/abs/1410.4502}{{\tt 1410.4502}}].

\bibitem{Hat16}
P.~W. {Hatfield}, S.~N. {Lindsay}, M.~J. {Jarvis}, B.~{H{\"a}u{\ss}ler},
  M.~{Vaccari} and A.~{Verma}, \emph{{The galaxy-halo connection in the VIDEO
  survey at 0.5<z<1.7}},
  \href{http://dx.doi.org/10.1093/mnras/stw769}{\emph{\mnras} {\bf 459} (July,
  2016) 2618--2631}, [\href{https://arxiv.org/abs/1511.05476}{{\tt
  1511.05476}}].

\bibitem{Zhe09}
Z.~{Zheng}, I.~{Zehavi}, D.~J. {Eisenstein}, D.~H. {Weinberg} and Y.~P. {Jing},
  \emph{{Halo Occupation Distribution Modeling of Clustering of Luminous Red
  Galaxies}}, \href{http://dx.doi.org/10.1088/0004-637X/707/1/554}{\emph{\apj}
  {\bf 707} (Dec., 2009) 554--572},
  [\href{https://arxiv.org/abs/0809.1868}{{\tt 0809.1868}}].

\bibitem{Mat11}
Y.~{Matsuoka}, S.~{Masaki}, K.~{Kawara} and N.~{Sugiyama}, \emph{{Halo
  occupation distribution of massive galaxies since z= 1}},
  \href{http://dx.doi.org/10.1111/j.1365-2966.2010.17464.x}{\emph{\mnras} {\bf
  410} (Jan., 2011) 548--558}, [\href{https://arxiv.org/abs/1008.0516}{{\tt
  1008.0516}}].

\bibitem{Zeh11}
I.~{Zehavi}, Z.~{Zheng}, D.~H. {Weinberg}, M.~R. {Blanton}, N.~A. {Bahcall},
  A.~A. {Berlind} et~al., \emph{{Galaxy Clustering in the Completed SDSS
  Redshift Survey: The Dependence on Color and Luminosity}},
  \href{http://dx.doi.org/10.1088/0004-637X/736/1/59}{\emph{\apj} {\bf 736}
  (July, 2011) 59}, [\href{https://arxiv.org/abs/1005.2413}{{\tt 1005.2413}}].

\bibitem{Puj17}
A.~{Pujol}, K.~{Hoffmann}, N.~{Jim{\'e}nez} and E.~{Gazta{\~n}aga}, \emph{{What
  determines large scale galaxy clustering: halo mass or local density?}},
  \href{http://dx.doi.org/10.1051/0004-6361/201629121}{\emph{\aap} {\bf 598}
  (Feb., 2017) A103}.

\bibitem{Lap12}
A.~{Lapi}, M.~{Negrello}, J.~{Gonz{\'a}lez-Nuevo}, Z.-Y. {Cai}, G.~{De Zotti}
  and L.~{Danese}, \emph{{Effective Models for Statistical Studies of
  Galaxy-scale Gravitational Lensing}},
  \href{http://dx.doi.org/10.1088/0004-637X/755/1/46}{\emph{\apj} {\bf 755}
  (Aug., 2012) 46}, [\href{https://arxiv.org/abs/1206.1142}{{\tt 1206.1142}}].

\bibitem{Cai13}
Z.-Y. {Cai}, A.~{Lapi}, J.-Q. {Xia}, G.~{De Zotti}, M.~{Negrello},
  C.~{Gruppioni} et~al., \emph{{A Hybrid Model for the Evolution of Galaxies
  and Active Galactic Nuclei in the Infrared}},
  \href{http://dx.doi.org/10.1088/0004-637X/768/1/21}{\emph{\apj} {\bf 768}
  (May, 2013) 21}, [\href{https://arxiv.org/abs/1303.2335}{{\tt 1303.2335}}].

\bibitem{topcat}
M.~B. {Taylor}, \emph{{TOPCAT \& STIL: Starlink Table/VOTable Processing
  Software}},  in \emph{Astronomical Data Analysis Software and Systems XIV}
  (P.~{Shopbell}, M.~{Britton} and R.~{Ebert}, eds.), vol.~347 of
  \emph{Astronomical Society of the Pacific Conference Series}, p.~29, Dec.,
  2005.

\bibitem{nedcc}
E.~L. {Wright}, \emph{{A Cosmology Calculator for the World Wide Web}},
  \href{http://dx.doi.org/10.1086/510102}{\emph{\pasp} {\bf 118} (Dec., 2006)
  1711--1715}, [\href{https://arxiv.org/abs/astro-ph/0609593}{{\tt
  astro-ph/0609593}}].

\bibitem{corner}
D.~Foreman-Mackey, \emph{corner.py: Scatterplot matrices in python},
  \href{http://dx.doi.org/10.21105/joss.00024}{\emph{The Journal of Open Source
  Software} {\bf 24} (2016) }.

\bibitem{astropy}
{Astropy Collaboration}, T.~P. {Robitaille}, E.~J. {Tollerud}, P.~{Greenfield},
  M.~{Droettboom}, E.~{Bray} et~al., \emph{{Astropy: A community Python package
  for astronomy}},
  \href{http://dx.doi.org/10.1051/0004-6361/201322068}{\emph{\aap} {\bf 558}
  (Oct., 2013) A33}, [\href{https://arxiv.org/abs/1307.6212}{{\tt 1307.6212}}].

\end{thebibliography}\endgroup
\bibliographystyle{JHEP.bst}



\end{document}